\documentclass[11pt,a4paper]{article}

\def\PWG{{\tt{POWHEG}}}
\usepackage{bold-extra}
\def\bPWG{{\tt{\textbf{POWHEG}}}}

\pdfoutput=1
\usepackage{color}
\usepackage{booktabs}
\usepackage{multirow,rotating}
\usepackage{amsmath,amssymb,bm,cite,graphicx,wasysym,mathrsfs,dsfont}
\usepackage{slashed,relsize}
\usepackage[colorlinks=true,linkcolor=blue,anchorcolor=black,citecolor=blue,filecolor=cyan,menucolor=red,runcolor=filecolor,urlcolor=blue,bookmarks=true,bookmarksnumbered=true]{hyperref}

\def\beq{\begin{equation}\displaystyle}
\def\eeq{\end{equation}}
\def\bea{\begin{eqnarray}\displaystyle} 
\def\eea{\end{eqnarray}}
\def\dst{\displaystyle}
\def\({\left(}
\def\){\right)}
\def\bry{\begin{array}}
\def\ery{\end{array}}
\def\l{\label}
\def\f{\frac}
\def\as{\alpha_{\rm\sc{s}}}

\usepackage{marginnote}

\def\W{\rm{W}}
\def\Y{\rm{Y}}

\title{\bf On the W\&{Y} interpretation of high-energy \\ Drell--Yan measurements}

\author{
Riccardo Torre$^{a,b}$, Lorenzo Ricci$^{c}$, Andrea Wulzer$^{b,c,d}$ \\ \\
{\small\emph{$^a$ INFN, Sezione di Genova, Via Dodecaneso 33, I-16146 Genova, Italy}}\\
{\small\emph{$^b$ CERN, 1211 Geneva 23, Switzerland}}\\
{\small\emph{$^c$ Theoretical Particle Physics Laboratory (LPTP), Institute of Physics,}}\\
{\small\emph{EPFL, Lausanne, Switzerland}}\\
{\small\emph{$^d$ Dipartimento di Fisica e Astronomia, Universit\`a di Padova, Italy}}\\
}
\begin{document}
\baselineskip=13pt
\begin{flushright}
CERN-TH-2020-180
\end{flushright}
\vspace{2em}
{\let\newpage\relax\maketitle}

\begin{abstract}
High-energy neutral and charged Drell--Yan differential cross-section measurements are powerful probes of quark-lepton contact interactions that produce growing-with-energy effects. This paper provides theoretical predictions of the new physics effects at the Next-to-Leading order in QCD and including one-loop EW corrections at the single-logarithm accuracy. The predictions are obtained from SM Monte Carlo simulations through analytic reweighting. This eliminates the need of performing a scan on the new physics parameter space, enabling the global exploration of all the relevant interactions. Furthermore, our strategy produces consistently showered events to be employed for a direct comparison of the new physics predictions with the data, or to validate the unfolding procedure than underlies the cross-section measurements. Two particularly relevant interactions, associated with the W and Y parameters of EW precision tests, are selected for illustration. Projections are presented for the sensitivity of the LHC and of the HL-LHC measurements. The impact on the sensitivity of several sources of uncertainties is quantified.
\end{abstract}

\newpage

\begingroup
\tableofcontents
\endgroup 

\setcounter{equation}{0}
\setcounter{footnote}{0}
\setcounter{page}{1}

\newpage
\section{Introduction}

Accurate measurements of high-energy observables are powerful probes of new physics, and arguably one of the most promising avenues for the continuation of the LHC experimental program. The study of neutral ($l^+l^-$) and charged ($l\nu$) Drell--Yan differential cross-section measurements offers a clear illustration of this potential~\cite{Farina:2016rws}, which however has also been demonstrated for several other processes, including diboson and boson-plus-Higgs \cite{Hagiwara:1986vm,Hagiwara:1989mx,Khachatryan:2015sga,Aad:2016ett,Butter:2016cvz,Zhang:2016zsp,Green:2016trm,Biekoetter:2014jwa,Falkowski:2015jaa,Baglio:2017bfe,Franceschini:2017xkh,Panico:2017frx,Grojean:2018dqj,Banerjee:2018bio,Liu:2018pkg,Henning:2018kys} and di-quark \cite{Alioli:2017nzr,Farina:2018lqo} production, and at future colliders (see ref.~\cite{Strategy:2019vxc} for a summary). The main goal of the present paper is to produce the theoretical tools needed to exploit the Drell--Yan (DY) measurements potential. While our results and methodologies are specific of the DY process, the challenges we face are of general nature. Some of the elements presented here could thus also be of help for the study of high-energy measurements in more complex final states.

The competitive advantage of high-energy measurements stems from the fact that the effects of heavy new physics, at a scale $\Lambda$, increase with the energy $E$ of the process as a (positive) power of $E/\Lambda$. Given finite measurements accuracy, new physics could thus be visible only at high enough $E$. The growing-with-energy behavior is easily understood by dimensional analysis in the Effective Field Theory (EFT) description of heavy new physics. Restricting as customary to dimension-$6$ operators, with Wilson coefficients $G\propto1/\Lambda^2$ of dimension $-2$, we immediately identify possible contributions to the scattering amplitudes of order $G\cdot E^2$ relative to the Standard Model. Such quadratically enhanced terms, and in turn the corresponding Wilson coefficients, are the target of high-energy measurements.

Only four fermion operators involving leptons and light quarks produce quadratically-growing tree-level terms in the DY amplitudes.\footnote{This holds at dimension-$6$, where in particular no gluon-lepton contact operator is present, and in an operators basis such as the Warsaw one \cite{Grzadkowski:2010es} where the operators $O_{2W}$ and $O_{2B}$ (defined in section~\ref{sec:rew}) are eliminated by the equations of motions in favor of four-fermions operators.} We can thus focus on these operators and ignore the others, whose non-enhanced effects are much smaller at high energy if their Wilson coefficients are comparable with the one of the enhanced (lepton-quark four fermion) operators. Moreover the non-enhanced operators are most likely probed more effectively in low-energy measurements, which are more precise than the high-energy ones due to the larger statistics. Therefore it would not be worth including the non-enhanced operators in the high-energy DY interpretation even if their Wilson coefficients were anomalously large. 

Among all possible lepton-quark operators (see~\cite{deBlas:2013qqa}) we can further restrict our attention to those of the ``current-current'' type, namely to the interactions of the form $J_{l}^\mu J_{q,\,\mu}$, with \mbox{$J^\mu_{\vphantom{l}}$\hspace{-7pt}$_{{l\,(q)}}$} any of the lepton (quark) chiral currents. This is because the fermion chirality structure of the other operators forbids them to interfere with the Standard Model (SM) amplitude.\footnote{Also Flavor-Changing Neutral Current (FCNCN) current-current operators, which are in any case irrelevant because of the strong flavor constraints (see however \cite{Fuentes-Martin:2020lea}), do not interfere with the SM.\label{fn1}} Therefore the ${\mathcal{O}}(G\,E^2)$ term they produce in the amplitude results in an ${\mathcal{O}}((G\,E^2)^2)$ contribution to the cross-section relative to the SM one. Since we are interested in probing new physics at scales $\Lambda$ that are higher than the available energy, so that $G\,E^2\ll1$, we can neglect non-interfering operators compared with the interfering (current-current) ones that do instead produce a genuine ${\mathcal{O}}(G\,E^2)$ contribution to the cross-section. The above argument of course fails, and non-interfering operators should be included, if their Wilson coefficients are enhanced relative to the current-current ones. However we are not aware of any concrete new physics scenario where this enhancement is structurally motivated, while it is easy to find scenarios where the converse happens and non-interfering operators are suppressed. Moreover when targeting quartically energy-growing effects (from the square of the new physics amplitude) one should also worry about the contribution of dimension-$8$ operators, that can produce similar effects at the interference level. We can thus regretless ignore non-interfering operators and focus on the current-current ones. This could of course be reconsidered at a more advanced stage of the global EFT interpretation of LHC data.

The methodology we develop in the present paper applies to any current-current operator involving light quarks (including the bottom) and leptons, with arbitrary current chirality and flavor structure. Concrete results are however only provided for two specific operators (i.e., $O_{2W}^\prime$ and $O_{2B}^\prime$, as defined in ref.~\cite{Farina:2016rws} and in the following section), whose Wilson coefficients correspond to the $\W$ and $\Y$ oblique parameters \cite{Barbieri:2004qk}. This choice is motivated in the first place by clarity, which we value at this stage more than completeness. Namely, the expected sensitivity and the impact of the various sources of uncertainties is much more effectively illustrated in a $2$-dimensional parameter space rather than in the large multidimensional space that corresponds to the entire set of current-current operators. Second, measuring (or bounding) the $\W$ and $\Y$ parameters in DY is sufficient for the global analysis of certain classes of EFTs. Namely for those of the ``universal''~\cite{Wells:2015uba} and (since the top quark plays no role here) ``top-philic'' \cite{Abramowicz:2018rjq,AguilarSaavedra:2018nen} type, that include motivated new physics scenarios such as Higgs (and Top) Compositeness~\cite{Giudice:2007fh}. Of course many more operators than $O_{2W}^\prime$ and $O_{2B}^\prime$ are present in these EFTs. However \mbox{$O^\prime$\hspace{-3pt}$_{2W/2B}$} are the only relevant ones in the DY process and thus the only ones to be included in this channel in view of a global fit. The ``W\&{Y} interpretation'' of DY measurements is thus an ideal (simple and informative) benchmark target for experimental analyses in this channel. Therefore we focus on these two operators, taking however into account that a larger set of current-current operators will have to be included at a second stage of the DY measurements interpretation. The reweighting strategy we develop in the paper will play in this extended analysis an even more vital role than what it does in the W\&{Y} case. We will return to this point later.

The search for EFT effects in DY data will be most likely based on unfolded differential cross-section measurements, similar to those in refs.~\cite{Aad:2016zzw,CMS:2014hga} and~\cite{CMS:2016pkm} for $8$~TeV and early run-$1$ data, to be compared with the corresponding EFT predictions.\footnote{The alternative is to compare the EFT predictions directly with the data distributions at the observed level. If this strategy is adopted, accurate Monte Carlo events generators are needed, and not only differential cross-section predictions. The reweighting strategy we propose does also provide accurate event samples.} We should then provide such predictions as accurately as possible and, equally importantly, provide reliable estimates of the associated parametrical and theoretical uncertainties. The target accuracy is dictated by the experimental error on the corresponding measurement, which is going to be vastly different in different energy regions. At very high energy the error will unavoidably get large, because of the limited statistics. This reduces the needs for theoretical accuracy, potentially allowing us to cope with the limited knowledge of Parton Distribution Functions (PDF), with the lack of Electroweak (EW) logs resummation, and with other effects that enhance the uncertainties at very high energy. Verifying to what extent this is indeed the case is one of the goals of the present paper. A lot of data are instead available at lower energy, and the measurement error will be dominated by systematic uncertainties. While we are unable to quantify them, based on refs.~\cite{Aad:2016zzw,CMS:2014hga} we expect experimental systematics of order few percent in the energy range from $300$~GeV to $2$~TeV. We will see if this accuracy goal can be met given state-of-the-art calculations and PDF uncertainties. 

It should be noted that the growing-with-energy nature of our signal is very well compatible with the hierarchical structure of the experimental and theoretical errors described above. At very high energy, where the error is larger, the signal is also larger, hence potentially visible. At lower energies the signal gets smaller, but still it is potentially visible because the error shrinks. We thus end up in a situation where the sensitivity to the signal comes from a wide range of energies, rather than from a single energy bin. This is ideal from two viewpoints. First, because we have the opportunity to improve the sensitivity by enlarging this range either on the high energy side, by collecting more luminosity, or on the low energy one by reducing the experimental systematic and theoretical uncertainties. Second, because the observation of a tension with the SM in multiple bins would be a very convincing evidence of new physics.

Current state-of-the-art EFT predictions for DY, at Next-to-Leading Order (NLO) in QCD and consistently interfaced with parton shower, are implemented in \PWG~\cite{Alioli:2018ljm}. The practical applicability of this tool however is limited by the fact that NLO simulations are long and demanding, and they should be run several times in order to extract, for each bin, the dependence of the cross-section on the EFT parameters. Of course the task is simplified by the fact that the cross-section is a quadratic polynomial in the Wilson coefficients. However extracting the polynomial coefficients (in particular, the linear ones) requires very accurate simulations, to be sensitive to the small correction due to the EFT on top of the SM. Moreover it requires a careful choice of the simulation parameters, which should be such that neither the SM nor the quadratic terms dominate by too many orders of magnitude. Since this parameters choice depends strongly on the bin, a large number of accurate simulations is required. While this approach might perhaps still work in the two-parameters W\&{Y} case, it would definitely be unfeasible in the large parameter space of generic current-current operators.

To solve the problem, in this paper we adopt a different methodology, based on event reweighting. Namely we notice that the Born, the virtual, and the real helicity amplitudes are all  affected by current-current operators through a common multiplicative factor. This factor is a linear polynomial in the Wilson coefficients (which enters squared in the cross-section), with constant term equal one corresponding to the SM contribution and coefficients that depend on the dilepton center-of-mass energy. The coefficients of the polynomial are readily computed for each combination of helicities and of quarks and lepton flavors, and they allow us to model the entire EFT parameter space, at exact NLO accuracy, by reweighting the events of a single Monte Carlo simulation. Namely, for each simulated event we compute the coefficients and we store them in the events file. Once a cross-section binning is defined, the events are binned accordingly and the stored information is used to compute the coefficients of the quadratic polynomial that describes the dependence on the EFT parameters of the cross-section in each bin. Reweighted  Monte Carlo events can also be used for the direct comparison of the EFT with the data, or in order to check the possible impact of the EFT effects on the unfolding procedure by which the cross-sections are measured. Notice that the obvious virtues of the reweighting methodology (whenever applicable) are well-recognized in the literature, to the point that reweighting has been automated in MadWeight~\cite{Artoisenet:2010cn}. However we cannot use MadWeight for our analysis because quark-lepton four-fermion operators are not yet included in the EFT {\sc{MadGraph}}~\cite{Alwall:2014hca} model at NLO ~\cite{Degrande:2020evl}. We produced our own implementation based on the SM \PWG~DY generator~\cite{Alioli:2008gx}.

Analytic reweighting is not only more efficient in providing state-of-the-art (NLO QCD) EFT predictions, but it also allows one to improve the accuracy of the predictions by including new effects. We consider in particular the EW single and double logs, which are enhanced at high mass and constitute the dominant NLO EW effects, and we include them in the EFT prediction. We also use reweighting to estimate the effect (on the SM prediction, most importantly) of Sudakov logs of higher orders in the loop expansion.

The paper is structured as follows. In section~\ref{sec:rew} we introduce our reweighting strategy and its implementation and show how to obtain EFT cross-section predictions and to estimate the corresponding uncertainties. As mentioned, our EFT predictions are accurate at the NLO in QCD and include NLO EW logs. Corrections due to QCD at NNLO and the complete NLO EW corrections can be straightforwardly added to the SM term using FEWZ~\cite{Li:2012wna} and \PWG~\cite{Barze:2012tt,Barze:2013fru}. In sections~\ref{DYL} and~\ref{sec:pro} we present LHC sensitivity projections, focusing on integrated luminosities of $100$~fb$^{-1}$, $300$~fb$^{-1}$, and $3$~ab$^{-1}$. These are obtained by a likelihood fit that includes all the relevant sources of theoretical uncertainties, allowing us to quantify their impact. Experimental systematic uncertainties are assumed at the few $\%$ level. We report our conclusions in section~\ref{sec:con}.

\newpage

\section{Reweighting strategy}\label{sec:rew}

We start, in section~\ref{fo}, by discussing how fixed-order QCD NLO predictions, in the presence of quark-lepton current-current new physics interactions, can be obtained by analytic reweighting. Next, in section~\ref{pwg}, we illustrate our \PWG~implementation and we show that reweighting is fully compatible with the \PWG~master formula, ensuring that showering effects are consistently included in our reweighted Monte Carlo events. We address in section~\ref{ewk} the slightly more technical problem of including EW logarithms of IR and UV (RG-running) nature.

\subsection{Fixed-order QCD corrections}\label{fo}

\begin{figure}[t]
\begin{center}
\includegraphics[width=.85\textwidth]{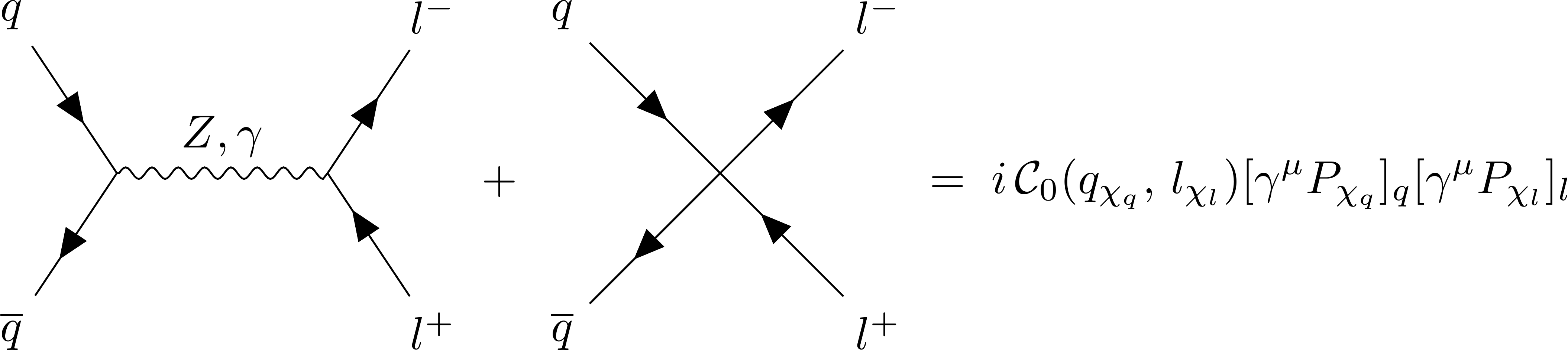}
\caption{\small Effective Feynman vertex for neutral DY, including SM EW and the new contact interactions.}\label{FR}
\end{center}
\end{figure}

We first consider neutral DY, i.e. the process
\beq
p\,p\,\rightarrow\,{{l}}^+{{l}}^-+X\,,
\eeq
with $l=e,\,\mu$ or (possibly) a $\tau$. We are interested in the high energy regime of the process, with a lower threshold on the dilepton center-of-mass energy that we set for definiteness at $\sqrt{s}>300$~GeV. In all the amplitudes that contribute to dilepton production, at the leading order in the EW and in the new physics couplings but at all orders in QCD, it is possible to isolate a common subdiagram, displayed in figure~\ref{FR} with its corresponding Feynman rule. In the figure, $\chi_q=L,\,R$ and $\chi_l=L,\,R$ denote the chirality of the quark and of the lepton legs, and $P_{\chi_{q,l}}$ the corresponding chirality projectors acting on the quarks and leptons spinor indices, respectively. Notice that only same-chirality $q$/${\overline{q}}$ and $l^+$/$l^-$ pairs can interact in the SM (Higgs interactions are of course totally negligible), and the same is true for the current-current effective vertices. Also the flavor ($q=u,d,s,c,b$) of the quark must be the same of the anti-quark since we are excluding FCNC new physics interactions as explained in footnote~\ref{fn1}. 

The effective coupling ${\mathcal{C}}_0$ depends on the quarks and leptons chirality and flavor, and it reads
\beq
{\mathcal{C}}_0(q_{\chi_q},\,l_{\chi_l})={\mathcal{C}}^0_{\rm{SM}}(s;q_{\chi_q},\,l_{\chi_l})+{\cal{K}}^0_{{{q_{\chi_q},\,l_{\chi_l}}}}\,,
\eeq
where ${\cal{K}}^0$ are constants that denote the coefficients of the effective neutral current interactions 
\beq
{\cal{K}}^0_{{q_{\chi_q},\,l_{\chi_l}}}({\overline{q}}_{\chi_q}\gamma^\mu q_{\chi_q})({\overline{l}}_{\chi_l}\gamma_\mu l_{\chi_l})\,.
\eeq
The SM contribution depends on the dilepton invariant mass and it can be concisely written as
\beq\label{SMC0}
{\mathcal{C}}^0_{\rm{SM}}(s;q_{\chi_q},\,l_{\chi_l})=\frac{g^2 T^3(q_{\chi_q}) T^3(l_{\chi_l}) + g^{\prime\,2} Y(q_{\chi_q}) Y(l_{\chi_l}) }{s-m_Z^2}+\frac{e^2Q(q)Q(l)m_Z^2}{s(m_Z^2-s)}\,,
\eeq
where $g$ and $g^\prime$ denote the SU$(2)_L$ and U$(1)_Y$ couplings, $e$ is the electric charge, $T^3$ is the third SU$(2)_L$ generator, $Y$ and $Q$ are the hypercharge and the fractional charge.\footnote{We follow the exact same conventions as in ref.~\cite{Grzadkowski:2010es}, apart from the sign (irrelevant in the above equation) of the coupling in the covariant derivatives and in the field-strengths definition.}

Based on the above discussion, it is obvious that at tree-level the dependence on the new physics parameters ${\mathcal{K}}^0$ can be obtained by reweighting the SM (${\mathcal{K}}^0=0$) predictions. The dilepton production cross-section (fully differential in the dilepton $4$-momenta) is the sum of the polarized $q{\overline{q}}\rightarrow l^+l^-$ partonic cross-sections convoluted with the corresponding PDF. The quarks and the leptons being effectively massless, each term in the sum depends on new physics through the square of the corresponding ${\mathcal{C}}_0(q_{\chi_q},\,l_{\chi_l})$ coefficient. The differential cross-section is thus the sum of the SM cross-sections in each helicity and quark flavor channel, each weighted by the factor
\beq\label{eqrew}
\rho_{{\rm{n}}}(s,\,{\mathcal{K}}^0;q_{\chi_q},\,l_{\chi_l})=
\left(\frac{{\mathcal{C}}_0(q_{\chi_q},\,l_{\chi_l})}{{\mathcal{C}}^0_{\rm{SM}}(s;q_{\chi_q},\,l_{\chi_l})}\right)^2
=\left(1+\frac{{\mathcal{K}}^0_{q_{\chi_q},\,l_{\chi_l}}}{{\mathcal{C}}^0_{\rm{SM}}(s;q_{\chi_q},\,l_{\chi_l})}\right)^2\,.
\eeq
Starting from a SM Monte Carlo simulation where quark and lepton flavors and helicities are stored in the events file (or, equivalently, from simulations of the individual channels), Monte Carlo events implementing the differential cross-section calculation are readily obtained by assigning to each event its reweighting factor $\rho$ as defined above. Of course we do not need to commit ourselves to a specific value of the ${\mathcal{K}}^0$ parameters and reweight the SM sample as the first step. Since $\rho$ is merely a linear polynomial (squared) in ${\mathcal{K}}^0$, with unit constant term, we just need to compute and store its coefficient $1/{{\mathcal{C}}^0_{\rm{SM}}}$ (plus the information of the helicity and flavor channel of the event) in the events file. The actual reweighting can be performed at a later stage, or one can use eq.~\eqref{eqrew} to compute the dependence on ${\mathcal{K}}^0$ of the cross-sections in the analysis bins.

The reweighting formula in eq.~\eqref{eqrew} also holds at NLO in QCD, because the gluon-quark coupling preserves the quark flavor and chirality. Therefore the one-loop $q_{\chi_q}{\overline{q}}_{\chi_q}\rightarrow l_{\chi_l}^+l_{\chi_l}^-$ amplitude is proportional to the same ${\mathcal{C}}_0(q_{\chi_q},\,l_{\chi_l})$ factor as the tree-level one, and the same is true for $q_{\chi_q}{\overline{q}}_{\chi_q}$--initiated real emission amplitudes with one final gluon and for $g\,q_{\chi_q}$-- and $g\,{\overline{q}}_{\chi_q}$--initiated real emissions. The dilepton differential cross-section is thus the linear combination, with $\rho$ reweighting coefficients as in eq.~\eqref{eqrew}, of Born plus virtual plus real contributions in each individual channel labeled by the flavor and chirality of the initial quark or anti-quark and by the ones of the leptons. Notice that the IR divergencies consistently cancel in each channel. The UV divergencies also cancel, with no renormalization needed for the new physics coupling because the new interaction involves a QCD-neutral vector current. Monte Carlo events reweighting can thus be carried out at NLO in the exact same way described above for the tree level case. It should be possible in principle to extend the reweighting approach also to NNLO accuracy. The main difference is that at NNLO new channels appear (like for instance the $ud\rightarrow ud\, l^+l^-$ real correction) whose amplitude is not proportional to one specific ${\mathcal{C}}_0(q_{\chi_q},\,l_{\chi_l})$ effective coupling, but to a linear combination of them. New reweighting factors should thus be computed and used to deal with these new channels. We do not explore this possibility because NLO accuracy will turn out to be more than sufficient for our purposes. NNLO corrections to the SM contribution are instead important, but those are easily added on top of our NLO new physics predictions.

\begin{figure}[t]
\begin{center}
\includegraphics[width=.75\textwidth]{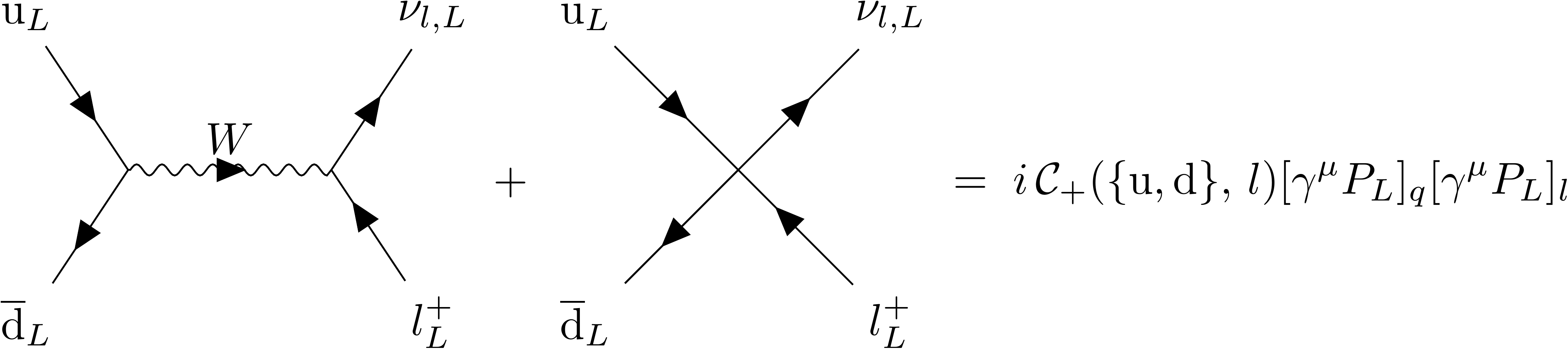}
\caption{\small Effective Feynman vertex for charge plus DY. The subscript ``$L$'' denotes chirality, not helicity.}\label{FRp}
\end{center}
\end{figure}

Analogous considerations hold for the charged DY process
\beq
p\,p\,\rightarrow\,{{l}}^\pm\, \overline{\nu}_l \hspace{-14.2pt}{\phantom l}^{\raisebox{.9pt}{\tiny \scalebox{.6}{$(\;\,\;\;\;)$}}}+X\,.
\eeq
The effective Feynman diagram is reported in figure~\ref{FRp} for charge-plus dilepton production, with ${\rm{u}}=u,c$ and ${\rm{d}}=d,s,b$ representing up- and down-type quark flavor indices. The effective coupling ${\mathcal{C}}_\pm(\{{\rm{u}},{\rm{d}}\},\,l)$ depends now on a pair of quark flavor indices denoted as ``$\{{\rm{u}},{\rm{d}}\}$'', and on the lepton flavor. It does not instead depend on the chirality, because all fermions are left-handed as indicated in the figure. This is obviously the case in the SM, but also for new physics  since the only relevant operator  \mbox{(i.e., {$O_{lq}$\hspace{-6pt}$^{(3)}$},} see table~\ref{operators}) is purely left-handed. We readily obtain the reweighting factors
\beq\label{eqrewch}
\rho_{\rm{c}}(s,\,{\mathcal{K}}^+;\{u,d\},\,l)=
\left|\frac{{\mathcal{C}}_+(\{{\rm{u}},{\rm{d}}\},\,l)}{{\mathcal{C}}^+_{\rm{SM}}(s;\{{\rm{u}},{\rm{d}}\},\,l))}\right|^2
=\left|1+\frac{{\cal{K}}^+_{\{{\rm{u}},{\rm{d}}\},\,l}}{{\mathcal{C}}^+_{\rm{SM}}(s;\{{\rm{u}},{\rm{d}}\},\,l)}\right|^2\,,
\eeq
for both charge plus and minus DY processes. Explicitly, the SM effective coupling reads
\beq
{\mathcal{C}}^+_{\rm{SM}}(s;\{{\rm{u}},{\rm{d}}\},\,l)=\frac{g^2}{2}\frac{V^*_{{\rm{u}}\,{\rm{d}}}}{s-m_W^2}\,,
\eeq
where $V$ is the CKM quark mixing matrix. New physics is encapsulated in the couplings of the effective charged current interactions 
\beq
{\cal{K}}^+_{\{{\rm{u}},{\rm{d}}\},\,l}({\overline{\rm{d}}}_L\gamma^\mu {\rm{u}}_L)({\overline{\nu}}_{l,L}\gamma_\mu l_L)\,+\,{\rm{h.c.}}\,.
\eeq

Charged DY NLO events reweighting can be performed, using eq.~\eqref{eqrewch}, with the exact same logic we described in the neutral case. Notice that we can regretless apply the charged reweighting factor to all the events in the simulation, in spite of the fact that it was derived for the Left-Left (LL) chirality subprocesses, because all the SM events are indeed of the LL type. The only (very minor) subtlety with charged DY reweighting is associated with real NLO corrections producing a top quark in the final state, through for instance the SM $b\,g\rightarrow t\,l^-{\overline{\nu}}_l$ subprocess. Given that the top is massive, and since we are excluding effective interactions involving the top quark, we cannot deal with this process with our strategy. However its contribution is totally negligible in the SM and we do not expect that new physics effects in the top sector could be large enough to make it detectable. Otherwise, the final states with an extra top quark could be isolated experimentally and studied separately.

 \begin{table}[t]
\begin{center}
\begin{tabular}{|l|}
\hline
\hspace{15pt}Generic current-current \\
\hline\\[-10pt]
$O_{lq}^{(3)}=(\bar{\rm{l}}_L\sigma_I \gamma^\mu {\rm{l}}_L)(\bar{\rm{q}}_L\sigma_I\gamma_\mu {\rm{q}}_L)\,,$\\[4pt]
$O_{lq}^{(1)}=(\bar{\rm{l}}_L\gamma^\mu {\rm{l}}_L) (\bar{\rm{q}}_L\gamma_\mu {\rm{q}}_L)\,,$\\[4pt]
$O_{eu}=(\bar{\rm{e}}_R\gamma^\mu {\rm{e}}_R) (\bar{\rm{u}}_R\gamma_\mu {\rm{u}}_R)\,,$\\[4pt]
$O_{ed}=(\bar{\rm{e}}_R\gamma^\mu {\rm{e}}_R) (\bar{\rm{d}}_R\gamma_\mu {\rm{d}}_R)\,,$\\[4pt]
$O_{lu}=(\bar{\rm{l}}_L \gamma^\mu {\rm{l}}_L) (\bar{\rm{u}}_R \gamma_\mu {\rm{u}}_R)\,,$\\[4pt]
$O_{ld}\hspace{0.11cm}=(\bar{\rm{l}}_L \gamma^\mu {\rm{l}}_L) (\bar{\rm{d}}_R \gamma_\mu {\rm{d}}_R)\,,$\\[4pt]
$O_{qe}=(\bar{\rm{q}}_L \gamma^\mu {\rm{q}}_L) (\bar{\rm{e}}_R \gamma_\mu {\rm{e}}_R)$\\[4pt]
\hline
\end{tabular}
\hspace{20pt}
\begin{tabular}{|l|}
\hline
\hspace{40pt}W\&{Y} current-current\\
\hline\\[-10pt]
$O_{2W}^\prime\hspace{-2pt}=J^{a,\mu}_LJ^a_{L,\mu}\,,\;\;\;J^{a,\mu}_L=
\sum\limits_{f}\overline{f}\gamma^\mu T^a f \,,$\\[4pt]
$O_{2B}^\prime\hspace{-0pt}=\hspace{-2pt} J^{\mu}_YJ_{Y,\mu}\,,\;\;\;\;\;\;J^{\mu}_Y=\sum\limits_f \overline{f}\gamma^\mu Y f
\,,$\\[4pt]
$G_{lq}^{(3)}=\frac12 G_{2W}^\prime\,,\;\;\;\;\; $\\[4pt]
$G_{lq}^{(1)}=-\frac1{6} G_{2B}^\prime\,,\;\;\; G_{eu}=-\frac4{3} G_{2B}^\prime\,,$\\[4pt]
$G_{ed}=\frac2{3} G_{2B}^\prime\,,\;\;\; \;\;\;\;\;G_{lu}=-\frac2{3} G_{2B}^\prime\,,$\\[4pt]
$G_{ld}=\frac1{3} G_{2B}^\prime\,,\;\;\; \;\;\;\;\;G_{qe}=-\frac1{3} G_{2B}^\prime$\\[4pt]
\hline
\end{tabular}
\end{center}
\caption{Left: Generic current-current operators in the notation of refs.~\cite{Grzadkowski:2010es,deBlas:2013qqa}. Right: The two operators associated with the W and Y parameters. Operator couplings are denoted with ``$G$''. \label{operators}}
\end{table}

We now specialize the general reweighting formulas to the subset of operators that are selected for the W\&{Y} interpretation. The W and Y parameters are defined in this paper as the coefficients of the four-fermion operators $O_{2W}^\prime$ and $O_{2B}^\prime$ reported in table~\ref{operators}. More precisely, we write
\beq\label{WYop}
G_{2W}^\prime=-\frac{g^2{\rm{W}}}{2 m_W^2}\,,\;\;\;\;\;G_{2B}^\prime=-\frac{g^{\prime\, 2}{\rm{Y}}}{2 m_W^2}\,.
\eeq
By performing a field redefinition (i.e., by using the equations of motion), $O_{2W}^\prime$ and $O_{2B}^\prime$ can be traded for the gauge/gauge operators $O_{2W}$ and $O_{2B}$ of ref.~\cite{Giudice:2007fh}. In turn, $O_{2W}$ and $O_{2B}$ generate ``oblique'' corrections to the $Z$ and photon propagators that can be encapsulated in the phenomenological parameters $\hat{\rm{W}}$ and $\hat{\rm{Y}}$ probed at LEP~\cite{Barbieri:2004qk}. The normalization is chosen in eq.~\eqref{WYop} such that $\hat{\rm{W}}=$W and $\hat{\rm{Y}}=$Y at tree-level. The relevance of $O_{2W}$ and $O_{2B}$ (and in turn of $O_{2W/2B}^\prime$) stems from the fact that they are the only dimension-$6$ operators that grow with the energy in DY to be generated by a new physics scenario where the light quarks and the leptons communicate with the new physics sector only through the SM gauge interactions. For more details, also on the correspondence between $O_{2W/2B}$ and $O_{2W/2B}^\prime$, see ref.~\cite{Farina:2016rws}.

By employing eq.~\eqref{WYop}, table~\ref{operators}, and the almost direct correspondence between the ${\cal{K}}^{0,+}$ couplings and the Warsaw basis operator coefficients, we immediately derive the neutral and charged DY reweighting factors for the W\&{Y} interpretation
\bea\label{rwfac}
&&\rho_{{\rm{n}}}(s,W,Y;q_{\chi_q},\,l_{\chi_l})= \left(1+a_W^{\rm{n}}(s;q_{\chi_q},\,l_{\chi_l})W+a_Y^{\rm{n}}(s;q_{\chi_q},\,l_{\chi_l})Y\right)^2\,,\nonumber\\
&&\rho_{{\rm{c}}}(s,W)= \left(1+a_W^{\rm{c}}(s) W\right)^2\,,
\eea
where the neutral and charged $a^{{\rm{n}},{\rm{c}}}$ coefficients are
\bea\label{rwcoeff}
&&a_W^{{\rm{n}}}(s;q_{\chi_q},\,l_{\chi_l}) = -\frac{g^2 T^3(q_{\chi_q}) T^3(l_{\chi_l})}{m_W^2 {\mathcal{C}}^0_{\rm{SM}}(q_{\chi_q},\,l_{\chi_l}) }\,,\;\;\;\;\;
a_Y^{{\rm{n}}}(s;q_{\chi_q},\,l_{\chi_l}) =- \frac{g^{\prime\,2} Y(q_{\chi_q}) Y(l_{\chi_l}) }{m_W^2 {\mathcal{C}}^0_{\rm{SM}}(q_{\chi_q},\,l_{\chi_l})}\,,\nonumber\\
&&a_W^{{\rm{c}}}(s) = -\frac{s-m_W^2}{m_W^2}\,,
\eea
with ${\mathcal{C}}^0_{\rm{SM}}$ as in eq.~\eqref{SMC0}. The neutral DY reweighting coefficients are independent of the lepton flavor and of the quark family, because $O^\prime$\hspace{-3pt}$_{2W/2B}$ are quark- and lepton-family independent. The $a_W^{{\rm{n}}}$ and $a_Y^{{\rm{n}}}$ coefficients can thus be computed for each SM Monte Carlo event based on the quark type ($\rm{u}$ or $\rm{d}$) and on the quark/lepton (LL, LR, RL or RR) chirality combinations, for a total of $8$ options. Actually $a_W^{{\rm{n}}}$ is non-vanishing only for LL-chirality events, which are thus the only ones bringing the dependence on W. This is because $O_{2W}$ can be viewed as a modification of the $W_3$/$W_3$ component, which only couples to left-handed fermions, of the neutral vector bosons propagators. By similar considerations it is easy to understand why the charged DY reweighting does not depend on Y ($O_{2B}$ does not affect the charged $W$-boson propagator) and why $a_W^{{\rm{c}}}$ is flavor-independent (i.e., the CKM factor drops). Notice that these features make reweighting for charged DY trivial, in the sense that all events have to be scaled with the same (but dependent on the dilepton mass) factor. Namely the charged dilepton differential cross-section is equal to the SM one, summed over all channels, times the overall factor $\rho_{{\rm{c}}}(s,W)$ that brings the entire dependence on new physics. This is of course not the case for neutral DY, where different flavor and helicity channels are weighted by different W\&{Y}-dependent factors.

\subsection[Reweighting \PWG]{Reweighting \bPWG}\label{pwg}

We applied our reweighting strategy to the \PWG~SM DY generator~\cite{Alioli:2008gx}. For the charged process, our procedure merely consists in computing and storing the reweighting coefficient $a_W^{\rm{c}}(s)$ in eq.~\eqref{rwcoeff} for each SM Monte Carlo event. The invariant mass $s=(p_l+p_\nu)^2$ is obtained from the lepton and neutrino momenta before the showering Monte Carlo ({\sc{Pythia}}~8~\cite{Sjostrand:2014zea}, in our case) is applied to the event. The augmented SM Monte Carlo sample can be used to produce histograms, with the cross-section (or, more generally, the total weight) of each bin obtained as the (positive or negative) SM weight of the event, times the reweighting factor \eqref{rwfac}, summed over the events that fall in the bin. Notice that the cross-section in the bin can be evaluated as a function of W. Namely one can expand $\rho_{\rm{c}}$ in W, evaluate the coefficient of the linear and of the quadratic term and sum them up (with the appropriate SM weights) separately over the events in the bin. We thus obtain the linear and quadratic coefficients of the polynomial that describes the cross-section in the bin as a function of W. The constant term of the polynomial is of course the SM prediction for the cross-section.

The procedure is only slightly more complicated in the neutral DY case, because subprocesses with different quark and lepton helicities must be reweighted with different factors, while the SM Monte Carlo collects them in a single (one for each quark flavor) unpolarized channel. This is not a problem for charged DY because the amplitudes are non-vanishing only for the LL polarization subprocess as previously explained. Therefore even if the Monte Carlo evaluates unpolarized cross-sections, the result is effectively the polarized one. Fortunately in the code implementing the SM neutral DY calculation of ref.~\cite{Alioli:2008gx} it is easy to access and modify the $Z$ and the photon chiral couplings to quarks and leptons. We can thus produce four SM generators, labeled as LL, LR, RL and RR, in which only the corresponding quark/lepton chiral couplings are present (and set to the SM value) while the others are set to zero. \PWG~evaluates the unpolarized cross-sections in each of the four cases, however the results are effectively polarized as discussed above for the charged process. The four Monte Carlo samples obtained by the four generators represent the contribution of the four helicity subprocesses, to be reweighted with the corresponding factor. For each event in each sample we compute $a_W^{\rm{n}}(s;q_{\chi_q},\,l_{\chi_l})$ and $a_Y^{\rm{n}}(s;q_{\chi_q},\,l_{\chi_l})$ as in eq.~\eqref{rwcoeff}, using the information of the quark flavor in the event. Finally we combine the four samples in the calculation of the cross-section as a function of W and Y similarly to what previously explained for the charged case.

The procedure outlined above is exact (in the limit of massless leptons and quarks) from the viewpoint of a fixed-order NLO QCD calculation. However \PWG~\cite{Frixione:2007vw} also describes the hardest parton showering emission, producing events that can be further showered without introducing double-counting. It is thus legitimate to ask if and how our procedure interferes with the \PWG~approach, possibly invalidating its consistency. In order to answer, we sketch below the implementation of the \PWG~method, in the presence of new physics, on each individual helicity subprocess. This is a trivial extension of refs.~\cite{Alioli:2008gx,Frixione:2007vw}, from which we borrow all notations.

The Born ($B$), the real ($R_{q{\overline{q}},g}$, $R_{g{\overline{q}},q}$ and $R_{qg,{\overline{q}}}$, summed/averaged over the gluon helicity) and the ``bare'' virtual ($V_b$) contributions, for given helicities, are equal to the appropriate $\rho$ factor times the corresponding SM expressions. The same applies to the bare factorization counterterms $G_{\oplus,b}$ and $G_{\ominus,b}$, since they emerge from the bare parton distribution functions in the tree-level term and are therefore proportional to $B$. The Catani--Seymour counterterms are also equal to $\rho$ times their SM expressions, again because they are proportional to the Born term. One might want to cross-check the latter statement because the Catani--Seymour formalism was developed~\cite{Catani:1996vz} to deal with unpolarized processes, while here we are considering a polarized one. The statement can be readily verified by direct calculation or by noticing that the Catani--Seymour formulas hold for a completely generic unpolarized process with arbitrary Born term, and that our polarized cross-sections are effectively the unpolarized cross-sections as computed in a theory where the $Z$ and the photon only couple to specific quark and lepton chiralities. The Catani-Seymour formulas must thus apply. A last potential subtlety is associated with the fact that the $\rho$ reweighting factor depends on the dilepton invariant mass $\sqrt{s}$ and that the Catani-Seymour counterterms are evaluated on an ``underlying-Born'' $2\rightarrow2$ kinematics that is obtained from the true $2\rightarrow3$ kinematics by a prescription that is, to some extent, arbitrary. Fortunately with the choice of ref.~\cite{Alioli:2008gx} the underlying Born dilepton invariant mass is identical to the true one, therefore the exact same $\rho(s)$ factor appears in the Born, virtual and real contributions and in all counterterms. The dilepton invariant mass is of course also consistently preserved in the reconstruction of the $3$-body kinematics out of the underlying Born $2$-body variables. 
\newpage

We conclude that the elements that appear in the \PWG~master formula (see eq.~(4.17) of ref.~\cite{Frixione:2007vw}), including the subtracted virtual and the real contribution decomposition in the two singular regions, all depend on new physics through the same $\rho(s)$ multiplicative factor. The same thus holds for the ${\overline{B}}$ term, which is a linear combination of the latter terms. The rescaling instead cancels in the Sudakov exponent, which contains the ``R/B'' ratio of real over Born, and in the real radiation term of the formula for the same reason. Consequently, the \PWG~master formula for the cross-section is also equal to $\rho(s)$ times the corresponding SM object. Each of the LL, LR, RL and RR generators described above implements the \PWG~formula  for the corresponding helicity subprocess with $\rho=1$ (i.e., in the SM), and in each of them the contributions from different quark flavors are treated separately. By reweighting based on the quark flavor of each event and joining the four helicity samples we thus obtain events that rigorously implement the \PWG~calculation of the Drell--Yan process in the presence of new physics. After passing them through a showering Monte Carlo program, these events consistently include showering effects at the NLO in QCD.

\begin{figure}[t]
\centering
\begin{minipage}{0.5\textwidth}
  \centering
  \includegraphics[width=1.02\linewidth]{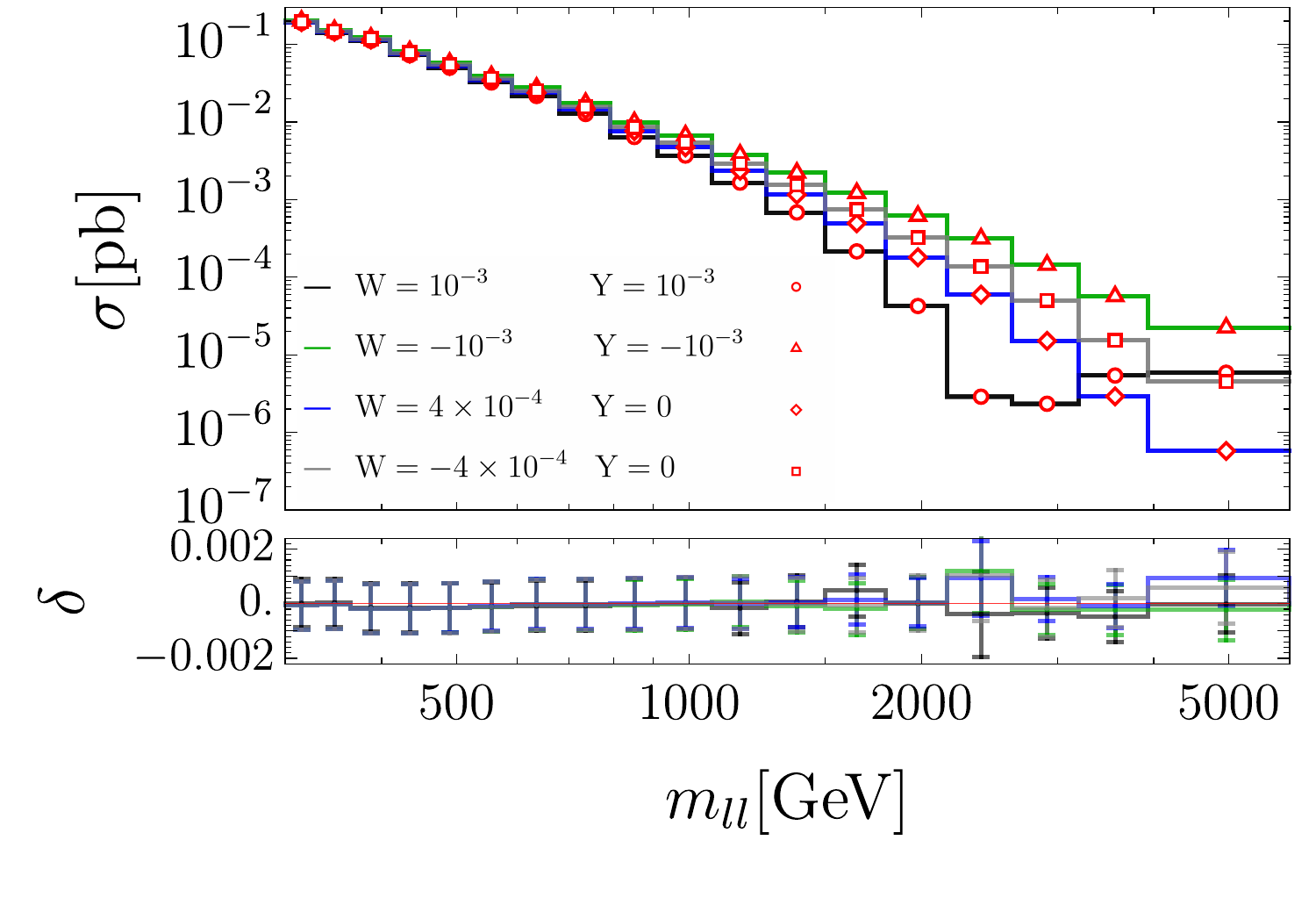}
\end{minipage}%
\begin{minipage}{0.5\textwidth}
  \centering
  \includegraphics[width=1.02\linewidth]{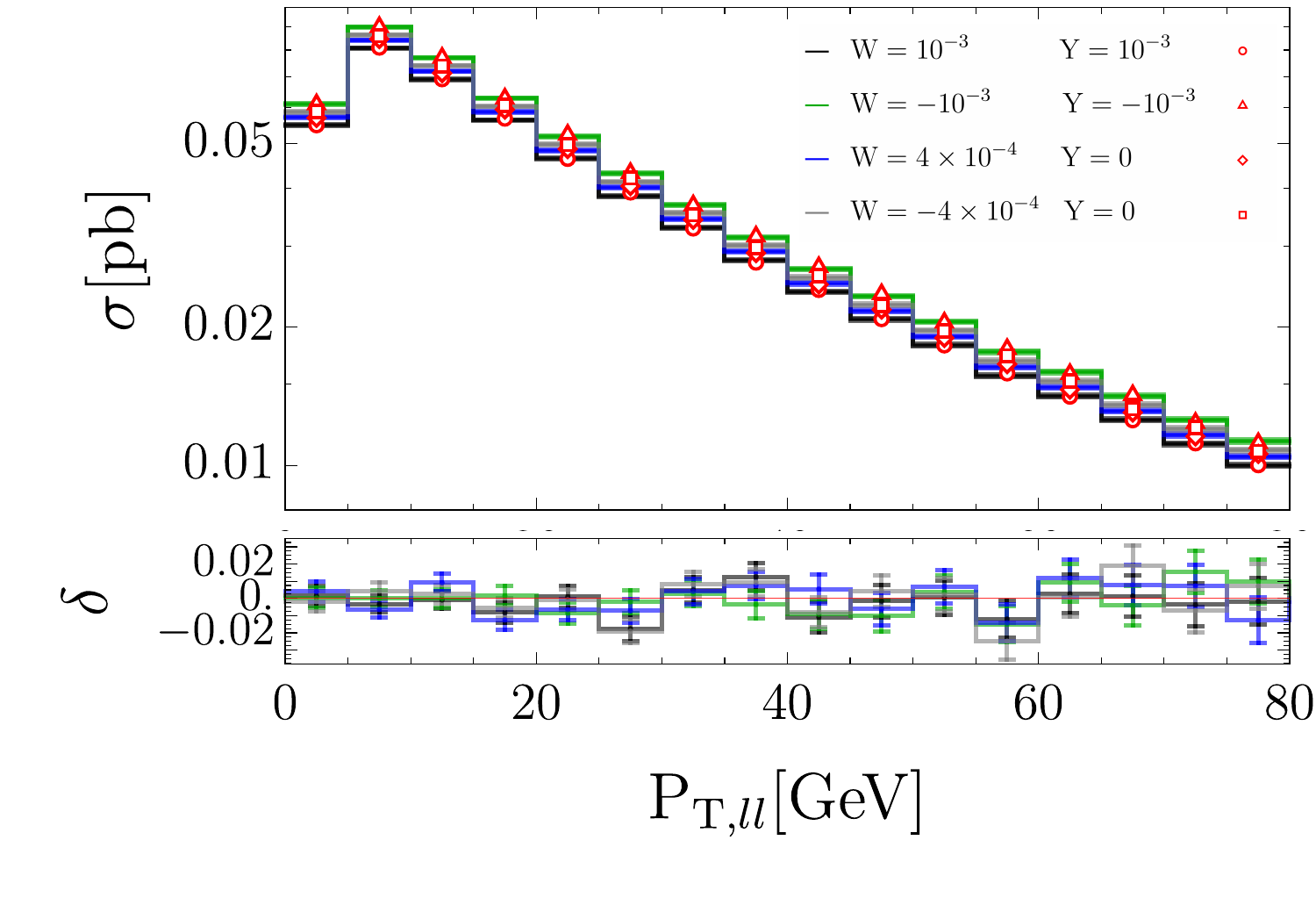}
\end{minipage}
\caption{Dilepton invariant mass ($m_{ll}$, left panel) and total transverse momentum of the dilepton pair ($\text{P}_{\text{T},ll}$, right panel) distributions. $\delta$ is the ratio between our prediction and the one of  ref.~\cite{Alioli:2018ljm} minus one, with its error obtained by combining the Monte Carlo errors of the predictions. \label{compfig}}
\end{figure}

Our reweighting strategy is one consistent implementation of the \PWG~method for the Drell--Yan calculation, but it is slightly different from the one of ref.~\cite{Alioli:2018ljm} (where SM EFT effects are included), and from the SM calculation~\cite{Alioli:2008gx}. This is because in these implementations, different helicity subprocesses are grouped into unpolarized channels as previously mentioned. Of course we eventually sum the four helicity contributions, but this is not sufficient to make our implementation identical to the other ones, because of the R/B ratio that appears in the Sudakov and in the real radiation term of the master formula. Our R/B is the ratio of real and Born terms where the external quarks and leptons have fixed helicity, while those of refs.~\cite{Alioli:2008gx,Alioli:2018ljm} are summed over the helicities. On inclusive observables the two implementations (after we sum over helicites, of course) give the same prediction at NLO, owing to the NLO accuracy of the \PWG~formula. The predictions are also identical at the leading log order where the real term in the Sudakov exponent and the real radiation term (for low-$k_T$ emissions) factorize as the product of the Born, which drops in the R/B ratio, times the appropriate splitting functions. Since the splitting functions are the same (notice that the gluon helicity sum is performed also in our case), the same expressions are found for R/B in the two implementations. The latter property clearly follows from the fact that the first \PWG~showering emission is consistent at the leading log level. The residual difference between the two implementations is thus beyond NLO and leading log accuracy, and too small to be appreciable in practice, as the results below demonstrate.

A validation of our reweighting is readily obtained by comparing with ref.~\cite{Alioli:2018ljm}, as in figure~\ref{compfig} and in table~\ref{comptable}. The left panel of the figure shows the neutral dilepton invariant distribution at four selected points in the W and Y parameter space as computed with our strategy, compared with those obtained with the code of ref.~\cite{Alioli:2018ljm}, represented as points. One minus the ratio between our prediction and the one of ref.~\cite{Alioli:2018ljm}, denoted as $\delta$, is displayed below the plot with the corresponding Monte Carlo error. A similar comparison is shown in table~\ref{comptable} for $9$ bins of the double-differential invariant mass and $\cos\theta_*$ (with $\theta_*$ the dilepton center-of-mass angle) distribution. The relative discrepancy $\delta$ is in all cases compatible with zero within the error. Notice that the error on $\delta$ is tiny in the invariant mass distribution plot because the cross-sections result from dedicated simulations in each bin. The error is larger in the doubly-differential distribution comparison because the cross-sections are obtained in this case by cutting the dedicated simulation events (of $10^5$ events each) in the $3$ $\cos\theta_*$ bins. In the right panel of the figure we consider instead the transverse momentum of the dilepton pair, integrated over the dilepton mass above $300$~GeV and over the angles. Although measuring this distribution is not relevant to probe W and Y, the comparison is interesting because of the slightly different implementation of the \PWG~radiation emission in the two approaches. Also in this distribution, no difference is found within the Monte Carlo error. Notice that the $\text{P}_{\text{T},ll}$ distribution includes showering with {\sc{Pythia}}~8~\cite{Sjostrand:2014zea}, while the other results described above are obtained with pure \PWG~events before showering. Other comparison plots were made, also for charged DY production, and no significant difference was found. 

\begin{table}[t]
\begin{center}
\begin{tabular}{|c|c|c|c|c|}
\cline{3-5}
\multicolumn{2}{c|}{} & \multicolumn{3}{c|}{$c_*$ bin} \\
\cline{3-5}
\multicolumn{2}{c|}{} & $[-1,-0.33]$ &  $[-0.33,0.33]$  &  $[0.33,1$]  \\
\hline
\multirow{3}{*}{\begin{sideways} $m_{ll}$ bin [GeV]\hspace{20pt} \end{sideways}} & $[330,365]$ & $\begin{array}{c}(-2\pm8)10^{-3}\\[-3pt] (-4\pm8)10^{-3}\\[-3pt] (-2\pm8)10^{-3}\\[-4pt] (0\pm8)10^{-3}\end{array}$ & $\begin{array}{c}(4\pm8)10^{-3}\\[-3pt] (3\pm8)10^{-3}\\[-3pt] (0\pm8)10^{-3}\\[-3pt] (4\pm8)10^{-3}\end{array}$  & 
$\begin{array}{c}(0\pm6)10^{-3}\\[-3pt] (0\pm6)10^{-3}\\[-3pt] (0\pm6)10^{-3}\\[-3pt] (0\pm6)10^{-3}\end{array}$  \\
\cline{2-5}
& $[910,1070]$ & $\begin{array}{c}(-6\pm8)10^{-3}\\[-3pt] (-12\pm8)10^{-3}\\[-3pt] (-15\pm8)10^{-3}\\[-3pt] (-6\pm8)10^{-3}\end{array}$ &
$\begin{array}{c}(7\pm8)10^{-3}\\[-3pt] (6\pm8)10^{-3}\\[-3pt] (0\pm8)10^{-3}\\[-3pt] (6\pm8)10^{-3}\end{array}$ 
 & $\begin{array}{c}(6\pm5)10^{-3}\\[-3pt] (8\pm5)10^{-3}\\[-3pt] (6\pm5)10^{-3}\\[-3pt] (6\pm5)10^{-3}\end{array}$  \\
\cline{2-5}
& $[2620,3200]$  & $\begin{array}{c}(-8\pm8)10^{-3}\\[-3pt] (3\pm8)10^{-3}\\[-3pt] (2\pm7)10^{-3}\\[-3pt] (-6\pm9)10^{-3}\end{array}$ & 
$\begin{array}{c}(2\pm7)10^{-3}\\[-3pt] (0\pm8)10^{-3}\\[-3pt] (-9\pm7)10^{-3}\\[-3pt] (8\pm8)10^{-3}\end{array}$ & 
$\begin{array}{c}(2\pm5)10^{-3}\\[-3pt] (-1\pm5)10^{-3}\\[-3pt] (4\pm5)10^{-3}\\[-3pt] (-2\pm6)10^{-3}\end{array}$  \\
\hline
\end{tabular}
\end{center}
\caption{Comparison with ref.~\cite{Alioli:2018ljm} in $9$ bins of the doubly-differential $m_{ll}$ and $c_*=\cos\theta_*$ distribution. The relative discrepancy $\delta$ is reported for the four values of the W\&{Y} parameter employed in figure~\ref{compfig}.}
\label{comptable}
\end{table}

\subsection{Electroweak logarithms}\label{ewk}

High-energy DY measurements target growing-with-energy new physics effects. Thus it is imperative to keep under control any SM contribution that might result in a similar behavior, such as EW double and single logarithms of both IR and UV (RG-running) origin. One-loop EW NLO corrections, including in particular the corresponding EW logs, are present in the neutral DY SM predictions of FEWZ~\cite{Li:2012wna} (together with QCD at NNLO), and in \PWG~both for charged and for neutral DY~\cite{Barze:2012tt,Barze:2013fru}. New physics must also be modeled correctly if we want to discover it by exploiting correlated deviations from the SM of the measurements in multiple bins. Clearly the new physics term is itself a correction to the SM, therefore it needs not to be predicted as accurately as the SM one. However one should still carefully monitor the impact of high-order corrections on the new physics contribution and include them if possible, as we did above for the NLO QCD corrections. We now show how to add, again through reweighting, EW logs at the one-loop order in the EW coupling expansion.

The relevant IR logs have been computed in ref.~\cite{Denner:2006jr} (see also refs.~\cite{Pozzorini:2001rs,Kuhn:1999nn,Kuhn:2001hz,Jantzen:2005az}) up to two loops, and they have been recently implement in Sherpa~\cite{Bothmann:2020sxm} (at one loop). Restricting to one loop, and defining 
\beq\label{logir}
L=\log\frac{s}{m_W^2}\,,\;\;\;\;\;
 L_{t (u)} =  2 L \log\frac{-t (u)}{s} +\log^2\frac{-t (u)}{s} \,,
\eeq
\ \\[-9pt]
\noindent the Feynman amplitudes for the fully exclusive $2\to2$ Drell--Yan processes at Next to Leading Logarithm (NLL) accuracy read \footnote{The equations that follow assume that the charge-minus amplitude is the conjugate of the charge-plus amplitude, as it is the case in the SM and for generic current-current New~Physics operators. The sum over the ${\rm{u}}^\prime$ and ${\rm{d}}^\prime$ flavor indices is understood. Log-enhanced terms with imaginary coefficient are not reported  because they do not interfere with the Born amplitudes.}
\begin{eqnarray}\label{IRLOG}
&&{\cal{M}}_{1{\rm{l}},{\rm{NLL}}}^{{\rm{u}}\overline{\rm{u}}\rightarrow l^- l^+}=
{\mathcal{F}}_{\rm{D}}
{\cal{M}}_{{\rm{B}}}^{{\rm{u}}\overline{\rm{u}}\rightarrow l^- l^+}
+\frac{g^2}{(4 \pi)^2} L_u\Re\left[
V_{{\rm{u}}{\rm{d}}^\prime}
{\cal{M}}_{{\rm{B}}}^{{\rm{u}}\overline{\rm{d}}^\prime\rightarrow  \nu_l l^+}
\right]\,,\\
&&
{\cal{M}}_{1{\rm{l}},{\rm{NLL}}}^{{\rm{d}}\overline{\rm{d}}\rightarrow l^- l^+}=
{\mathcal{F}}_{\rm{D}}
{\cal{M}}_{{\rm{B}}}^{{\rm{d}}\overline{\rm{d}}\rightarrow l^- l^+}
-\frac{g^2}{(4 \pi)^2} L_t \Re\left[
{\cal{M}}_{{\rm{B}}}^{{\rm{u}^\prime}\overline{\rm{d}}\rightarrow \nu_l l^+}V_{{\rm{u}^{\prime}}{\rm{d}}}
\right]\,,\nonumber\\
&&
{\cal{M}}_{1{\rm{l}},{\rm{NLL}}}^{{\rm{u}}\overline{\rm{d}}\rightarrow  \nu_l l^+}=
{\mathcal{F}}_{\rm{D}}
{\cal{M}}_{{\rm{B}}}^{{\rm{u}}\overline{\rm{d}}\rightarrow \nu_l l^+}
+\frac{g^2}{2(4\pi)^2}L_u
\left({\cal{M}}_{{\rm{B}}}^{{\rm{u}}\overline{\rm{u}}^\prime\rightarrow l^- l^+}V_{{\rm{u}}^\prime{\rm{d}}}^*+V_{{\rm{u}}{\rm{d}^\prime}}^*{\cal{M}}_{{\rm{B}}}^{{\rm{d}^\prime}\overline{\rm{d}}\rightarrow \nu_l \overline{\nu}_l } \right)\nonumber\\
&&
\hspace{127pt}-\;\frac{g^2}{2(4 \pi)^2}L_t\left(V_{{\rm{u}}{\rm{d}^\prime}}^* {\cal{M}}_{{\rm{B}}}^{{\rm{d}^\prime}\overline{\rm{d}}\rightarrow l^- l^+}+{\cal{M}}_{{\rm{B}}}^{{\rm{u}}\overline{\rm{u}}^\prime\rightarrow \nu_l \overline{\nu}_l} V_{{\rm{u}^\prime}{\rm{d}}}^* \right)\,.\nonumber
\end{eqnarray}
In the equation, ${\cal{M}}_{{\rm{B}}}$ denote the Born (tree-level) amplitudes, including their dependence on new physics encapsulated in the ${\mathcal{C}}_0$ and ${\mathcal{C}}_\pm=({\mathcal{C}}_\mp)^*$ effective couplings defined in section~\ref{fo}. The charged process amplitudes are of course only non-vanishing for the LL chirality process. Neutral amplitudes for ${\rm{u}}\overline{\rm{u}}^\prime\rightarrow \nu_l \overline{\nu}_l$ are equal to those for ${\rm{d}}\overline{\rm{d}}^\prime\rightarrow l^- l^+$ and similarly for down-initiated neutrino production. We denote as $q_1$, ${\overline{q}}_2$, $l_1$ and ${\overline{l}}_2$ the four particles involved in the scattering with the corresponding chiralities, such that the generic Drell--Yan partonic subprocess is
\beq\label{not}
q_1{\overline{q}}_2\rightarrow l_1 {\overline{l}}_2\,.
\eeq
With this notation, the Mandelstam variables are defined as 
\beq
s=(p_{q_1}+p_{{\overline{q}}_2})^2\,,\;\;\;\;\;t=(p_{q_1}-p_{l_1})^2\,,\;\;\;\;\;u=(p_{q_1}-p_{{\overline{l}}_2})^2\,.
\eeq

The ``diagonal'' ${\mathcal{F}}_{\rm{D}}$ factors in eq.~\eqref{IRLOG} depend on the fermion species and chiralities. They contain angular independent (a.i.) and angular dependent (a.d.) contributions. The latter ones emerge, together with the other angular-dependent terms in eq.~\eqref{IRLOG} (those proportional to $L_t$ and $L_u$), from the double logarithms of $t/m_W^2$ and of $u/m_W^2$ rewritten in terms of $L=\log{s/m_W^2}$. The a.d.~contributions which are not proportional to $L$ (e.g., the last term in eq.~\eqref{logir}), are normally not retained at the NLL accuracy. We do include them because they are enhanced in the forward and backward regions. We have verified that they considerably improve the quality of the NLL approximation, not only in the angular but also in the invariant mass dilepton distribution. 

We write ${\mathcal{F}}_{\rm{D}}$ as
\beq\label{FD}
{\mathcal{F}}_{\rm{D}}=f_{\rm{a.i.}} + f_{\rm{a.d.}}(t/s,u/s)+f_{\rm{a.i.}}^{\rm{qed}}+ f_{\rm{a.d.}}^{\rm{qed}}(t/s,u/s)\,,
\eeq
where the (IR-divergent) angular-independent and angular-dependent contributions from soft and collinear photon loops have been isolated in the corresponding $f^{\rm{qed}}$ terms, to be discussed later. The others can be written concisely as
\bea\label{f}
&&f_{\rm{a.i.}}=\frac{1}{2 (4 \pi)^2}
\hspace{-5pt}
\sum_{f=q_{1,2},l_{1,2}}
\hspace{-4pt}
\left[(-L^2+3L)\left( g^2 {\rm{C}}_f + g^{\prime\,2} y_f^2 - e^2q_f^2 \right)+2Lg_{Z,f}^2 \log\frac{m_Z^2}{m_W^2} \right]\,,
\\
&&f_{\rm{a.d.}}=\frac{1}{(4 \pi)^2} \left[(g_{Z,q_1} g_{Z,l_2}+g_{Z,q_2} g_{Z,l_1}  )\left(L_u+2\log(-u/s)\log\frac{m_Z^2}{m_W^2}\right)\right.\nonumber\\
&&\qquad\qquad\;\;\;\;\;\;\left.
-(g_{Z,q_1} g_{Z,l_1}+g_{Z,q_2} g_{Z,l_2}  )\left(L_t+2\log(-t/s)\log\frac{m_Z^2}{m_W^2}\right) \vphantom{\frac12}\right]  \,,\nonumber
\eea
in terms of the $T^3$ eigenvalue ($t^3_f$), the Casimir (${\rm{C}}_f=0,3/4$), the charge and hypercharge \mbox{($y_f$ and $q_f$)} of each of the four fermions $f=q_1,q_2,l_1,l_2$. The coupling of the $Z$ boson $g_{Z,f}= g (t^3_f - s_{\rm{w}}^2 q_f)/{c_{\rm{w}}}$ is used in place of $t^3_f$ for more compact expressions.

The results above are in $D=4-2\,\epsilon$ dimensions and the UV singularities are subtracted in the $\overline{\rm{MS}}$ renormalization scheme. The photon and the fermions are exactly massless, therefore soft and collinear divergences appear in the $f^{\rm{qed}}$ terms
\bea
&&f_{\rm{a.i.}}^{\rm{qed}}=-\frac{e^2}{2 (4 \pi)^2}\left(\frac2{\epsilon^2}+\frac3{\epsilon}\right)
\hspace{-5pt}
\sum_{f=q_{1,2},l_{1,2}}
\hspace{-4pt}
q_f^2\,,\\
&&f_{\rm{a.d.}}^{\rm{qed}}=\frac{2e^2}{(4 \pi)^2}\frac1\epsilon
\left[(q_{q_1}q_{l_1}+q_{q_2}q_{l_2})\log(-u/s)
-(q_{q_1}q_{l_2}+q_{q_2}q_{l_1})\log(-t/s)
\vphantom{\frac12}\right]
\,.\nonumber
\eea
Notice that the $f^{\rm{qed}}$ terms diverge, but they do not depend on $L=\log(s/m_W^2)$. This is because they are defined as the contribution to the loop integrals from the region where the virtual photon is soft and/or collinear to an external leg, and these regions are insensitive to $m_{W/Z}$ up to $m_{W/Z}^2/s$ power corrections. 

The $\epsilon$ poles get canceled by real corrections and by PDF renormalization in the calculation of the dilepton differential cross-section, provided extra emissions are allowed and provided the charged leptons momenta are defined by recombining collinear photons. If the energy (or $p_T$) threshold for extra photons, the $k_T^{\rm{rec}}$ transverse momentum threshold for recombination, and the factorization scale are a considerable fraction of $\sqrt{s}$, no large finite contributions emerge from the cancellation and the $f^{\rm{qed}}$ terms can simply be dropped in the cross-section calculation. We construct our reweighted Monte Carlo samples targeting the ``fully-inclusive'' differential cross-section as defined above. More exclusive results, incorporating in particular the effect of a lower (or absent) $k_T^{\rm{rec}}$ threshold (or of a small $\Delta{R}^{\rm{rec}}$ recombination cone), are easily obtained by passing the events through a QED showering Monte Carlo code. 

Reweighted Monte Carlo samples implementing the calculation described above are easily obtained from a LO generator, which employs the SM Born matrix element ${\cal{M}}_{{\rm{B}},{\rm{SM}}}$. Provided of course that the fermion chirality channels are treated separately, or based on LL, LR, RL, and RR polarized generators constructed as in section~\ref{pwg}, one can compute the reweighting factor
\bea\label{rwlog}
&&\rho^{q_1{\overline{q}}_2\rightarrow l_1 {\overline{l}}_2}_{\rm{NLL}}(s,t,u)=
\frac{|{\cal{M}}_{{\rm{B}}}^{q_1{\overline{q}}_2\rightarrow l_1 {\overline{l}}_2}|^2}{|{\cal{M}}_{{\rm{B}},\,{\rm{SM}}}^{q_1{\overline{q}}_2\rightarrow l_1 {\overline{l}}_2}|^2}
+ \frac{2\,
\Re\left[
{\cal{M}}_{1{\rm{l}},{\rm{NLL}}}^{q_1{\overline{q}}_2\rightarrow l_1 {\overline{l}}_2}({\cal{M}}_{{\rm{B}}}^{q_1{\overline{q}}_2\rightarrow l_1 {\overline{l}}_2})^*
\right]}{|{\cal{M}}_{{\rm{B}},\,{\rm{SM}}}^{q_1{\overline{q}}_2\rightarrow l_1 {\overline{l}}_2}|^2}\nonumber\\
&&\qquad\qquad\qquad\;\;\;\;\;\equiv\rho_{{\rm{n}}({\rm{c}})}^{q_1{\overline{q}}_2\rightarrow l_1 {\overline{l}}_2}+\Delta\rho^{q_1{\overline{q}}_2\rightarrow l_1 {\overline{l}}_2}_{\rm{NLL}}
\,,
\eea
for each event, as a function of the new physics couplings, and use it in a way similar to that described in the previous sections for the NLO QCD reweighting. Up to running effects, to be discussed below, the first term on the first line of the equation coincides with $\rho_{{\rm{n}},{\rm{c}}}$ in eqs.~\eqref{eqrew} and \eqref{eqrewch} for the neutral and charged processes, respectively. Namely, it can be expressed in terms of the ${\mathcal{C}}_0$ and ${\mathcal{C}}_\pm$ neutral and charged amplitude coefficients and the corresponding SM expressions. It is a perfect square and, restricting to the W\&{Y} case for concreteness, its dependence on new physics can be parametrized by the reweighting coefficients $a_{W/Y}^{\rm{n}}$ and $a_W^{\rm{c}}$ as in eq.~\eqref{rwfac}. The second term in eq.~\eqref{rwlog} contains NLL effects. It can also be expressed in terms of the $\mathcal{C}$'s using eq.~\eqref{IRLOG} and noticing that the spinor current matrix elements drop in the amplitude ratio. The $\Delta\rho_{\rm{NLL}}$ term is still a quadratic polynomial in the new physics parameters, but it is not a perfect square and its constant term is not equal to zero. This is due to the fact that NLL corrections are introduced also on the SM term, therefore the reweighting is non-trivial even in the absence of new physics. The $6$ coefficients of the $\Delta\rho_{\rm{NLL}}$  polynomial have to be computed and stored in the events file, together with $a_{W/Y}^{\rm{n}}$ and $a_W^{\rm{c}}$, in order to obtain the analytical W\&{Y}~predictions at NLL~EW accuracy. 

Notice that the NLL corrections are often negative, so that $\rho$ can become negative in certain regions of the phase space. If this had to result in a negative cross-section after the weights are summed up in some bin, it would mean that the EW IR corrections are too large to be treated perturbatively in that bin. Fortunately this does not occur in the energy range of interest for the LHC measurements. Finally, we remark that the NLL corrections in the LL chirality channel make neutral contact interaction operators contribute to the charged DY process, because of the amplitude mixing in eq.~\eqref{IRLOG}. Therefore, at least in principle, charged DY measurements are actually also sensitive to the Y parameter and not only to W.

So far we only discussed EW logs of IR origin. UV logs do not appear explicitly in eq.~\eqref{IRLOG} because we are applying the results of ref.~\cite{Denner:2006jr} with the $\overline{\rm{MS}}$ renormalization scale set to the center-of-mass energy $\sqrt{s}$. At one loop order this is irrelevant for the one-loop corrections in eq.~\eqref{IRLOG}, which can still be computed at a fixed scale. However the tree-level amplitudes need to be evaluated with running couplings, RG-evolved at the scale $\sqrt{s}$. When expanding at one-loop, this produces single logarithms of $s$. The running of the SM couplings $g$ and $g^\prime$ starts at the $Z$-boson mass $m_Z\simeq m_W$, where these parameters are defined. Therefore the SM couplings renormalization produces ``IR-type'' logarithms of $s/m_{W}^2$. These are readily computed by replacing
\beq\label{Running}
g^2\rightarrow g^2(s)\simeq g^2+\delta g^2= g^2 + \frac{g^4}{16\pi^2}b_g L\,,\;\;\;\;\;g^{\prime\,2}\rightarrow g^{\prime\,2}(s)\simeq g^{\prime\,2}+\delta g^{\prime\,2}= g^{\prime\,2} + \frac{g^{\prime\,4}}{16\pi^2}b_{g^\prime} L\,,
\eeq
in the ${\mathcal{C}}^{\pm,0}_{\rm{SM}}$ effective couplings as they appear in the effective Feynman vertices in figures~\ref{FR} and~\ref{FRp}. The factors $b_g= -{19}/{6}$ and $b_{g^\prime}= {41}/{6}$ in the above equation are the SM $g$ and $g^\prime$ couplings $\beta$-functions. 

The RG running of the new physics couplings is the last source of enhanced logarithms. However these are not logarithms of $s/m_W^2$, but rather of $\Lambda^2/s$, where $\Lambda$ is the scale where the EFT operators are renormalized. The explicit form of these terms depends on the definition of the renormalized W and Y parameters. These are given by eq.~\eqref{WYop}, in terms of the $G_{2W/2B}^\prime$ four-fermion operator coefficients renormalized at $\Lambda$. The SM parameters ($g$, $g^{\prime}$, and $m_W$) that appear in the equation are evaluated at $m_Z$, therefore they do not contribute to the running. Insertions of $O_{2W/2B}^\prime$ in EW loops generates RG-running logarithmic contributions to a number of dimension-six operators. However, only the current-current quark-lepton non-FCNC operators listed in table~\ref{operators} produce quadratically energy-growing effects in the DY cross-section and need to be retained. The effect of the others is power-suppressed relative to the leading energy-growing terms. It should be noted that $O_{2W/2B}^\prime$ generate generic quark-lepton operators, namely the universality relations on the right panel of table~\ref{operators} are violated by RG-running. In order to include running effect we thus need to go back to eqs.~\eqref{eqrew} and \eqref{eqrewch}, evaluated with the ${\mathcal{K}}$'s obtained by solving the evolution equations at the leading log. The final expression for the reweighting factors takes the form 
\beq\label{rwgfull}
\rho^{q_1{\overline{q}}_2\rightarrow l_1 {\overline{l}}_2}_{\rm{NLL}}(s,t,u)\equiv\rho_{{\rm{n}}({\rm{c}}),\Lambda}^{q_1{\overline{q}}_2\rightarrow l_1 {\overline{l}}_2}+\Delta\rho^{q_1{\overline{q}}_2\rightarrow l_1 {\overline{l}}_2}_{\rm{NLL},\Lambda}
\,.
\eeq
Explicit results, obtained with the {\tt{DsixTools}}~\cite{Celis:2017hod} calculation of the relevant $\beta$-functions for the new physics couplings, and including the running of the SM couplings, are presented in appendix~\ref{RW}.

The RG EFT logs are found to have a marginal impact on the phenomenological analysis of the DY data, but they introduce conceptually novel aspects that is worth clarifying. First, they introduce a dependence on the EFT operators renormalization scale $\Lambda$. Technically, $\Lambda$ is arbitrary and we conventionally set it to $\Lambda=10$~TeV in our projections for the W and Y parameters sensitivity. On the other hand, for the interpretation of the results in the microscopic UV theory the EFT operators emerge from, setting $\Lambda$ to the cutoff scale of the EFT would have been preferable. The EFT cutoff intrinsically depends on the UV theory. The choice $\Lambda=10$~TeV corresponds to the estimated cutoff scale in the Composite Higgs UV scenario with moderate $g_*$, for values of W and Y close to the LHC reach~\cite{Farina:2016rws}. Naively, one could consider employing the results with $\Lambda=10$~TeV also for EFT's with much higher cutoff, by running the operator coefficients down to $10$~TeV. However this would not be correct in general because running produces many operators at $10$~TeV, while our calculation assumes that $O_{2W/2B}^\prime$ are the only non-vanishing current-current operators at $\Lambda=10$~TeV. Therefore our results are strictly speaking inapplicable even to theories where $O_{2W/2B}^\prime$ are the only operators that emerge at the cutoff scale, if the cutoff scale is much higher than $10$~TeV. Furthermore even in theories with $10$~TeV cutoff, the presence of other operators, even if not of the current-current type, does influence the current-current operators running below $\Lambda$ and our calculation does not apply. While of limited practical relevance (since the RG logs are very small and the cutoff is unlikely to be much higher than $10$~TeV), this issue could be readily addressed by including in the reweighting all the current-current quark-lepton non-FCNC operators, with coefficients RG-evolved starting from the most general $d=6$ operators content at the scale $\Lambda$. Since current-current operators are the only relevant ones in DY up to power-suppressed effects, this will produce complete NLL predictions in the general EFT parameter space.

\begin{figure}[t]
\centering
\begin{minipage}{0.485\textwidth}
  \centering
  \includegraphics[width=1\linewidth]{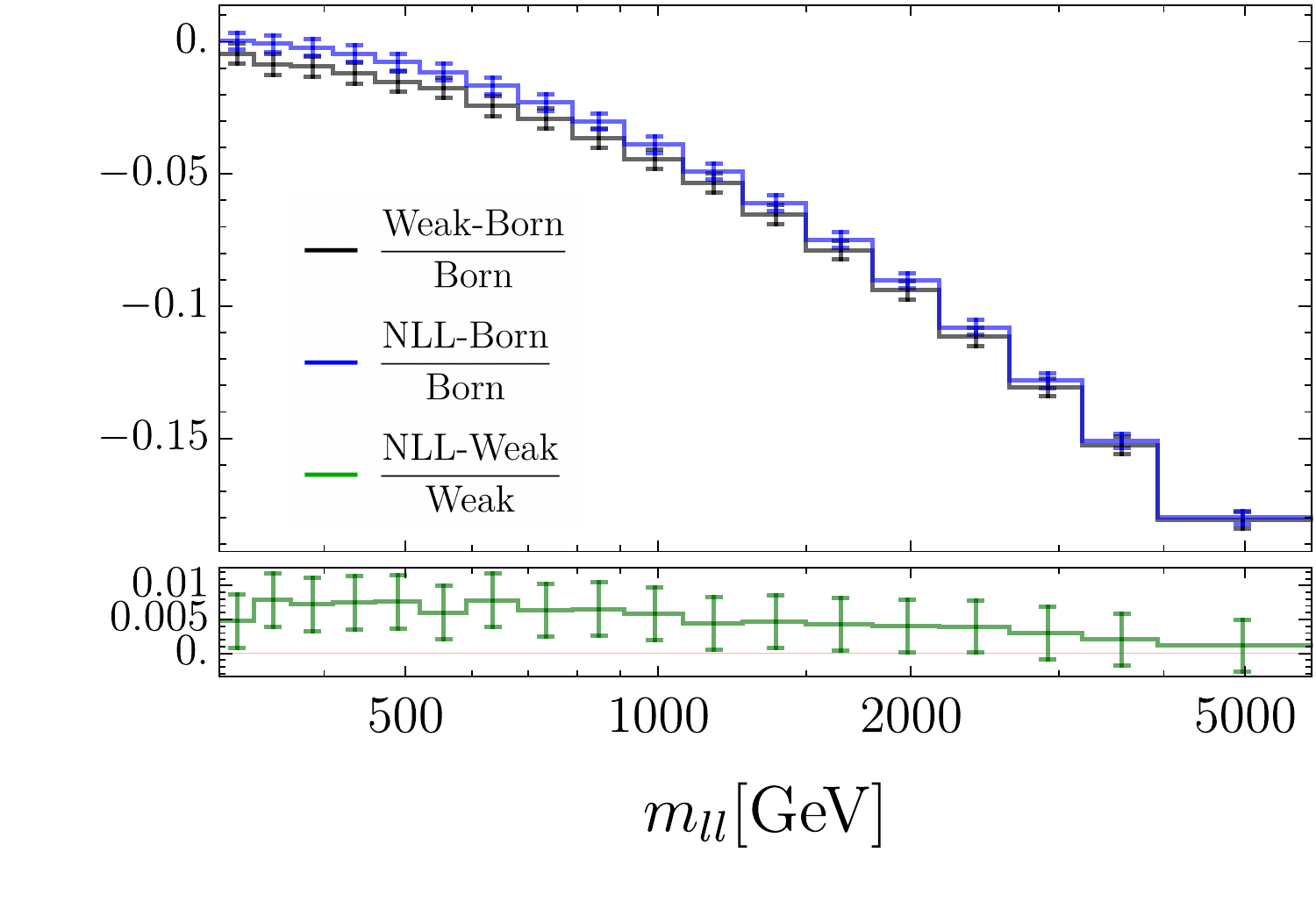}
\end{minipage}\hfill
\begin{minipage}{0.485\textwidth}
  \centering
  \includegraphics[width=1\linewidth,trim=10 125 0 0]{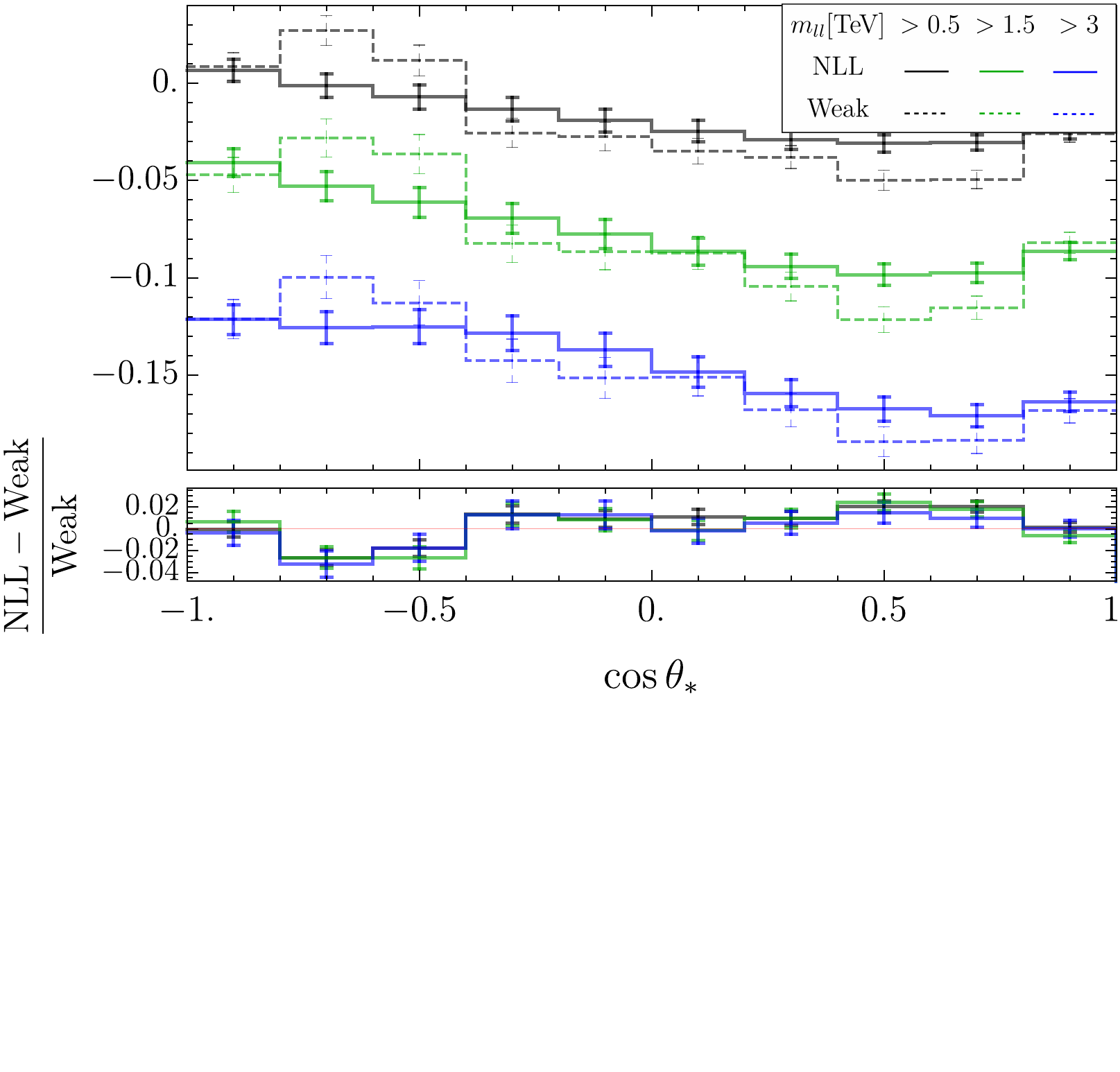}
\end{minipage}
\caption{Dilepton invariant mass ($m_{ll}$, left panel) and canter of mass angle ($\cos\theta_*$, right panel) distributions. $\delta$ is the discrepancy relative to the Born. \label{compWeak}}
\end{figure}

We can not fully validate our implementation of the EW logarithms because EW radiative corrections in the presence of the EFT operators have not been computed. However we can validate the SM EW logarithms using the \PWG~calculation of neutral DY including the complete one-loop EW correction~\cite{Barze:2013fru}. We employ the ``weak-only'' \PWG~ routine, that implements only the virtual corrections involving massive vector bosons, obtaining excellent agreement as figure~\ref{compWeak} shows. The logarithms reproduce the exact one-loop result up to ${\mathcal{O}}(1\%)$ accuracy, relative to the tree-level, in the entire mass spectrum. In particular they reproduce very accurately the ${\mathcal{O}}(20\%)$ enhancement of the corrections in the very high mass tail. Somewhat larger discrepancies are found as expected in the forward and backward regions of the angular distribution. As previously mentioned, the agreement would significantly deteriorate if we had not fully retained the angular-dependent logs. Given the expected statistics and experimental errors, $1\%$ accuracy in the predictions would probably be sufficient in the analysis, therefore one might even consider using the reweighted Monte Carlo in place of \PWG~for the Standard Model prediction. The same level of accuracy is expected in the prediction of the new physics EFT effects, relative to the exact one-loop calculation. Since new physics is itself a small correction to the SM (never more than $10\%$ in the relevant configurations), the reweighted prediction of the new physics term is fully equivalent to the exact one-loop result to all practical purposes. A technical aspect worth mentioning is that, since the ``weak-only''  \PWG~routine does not implement the box diagrams involving the exchange of a photon and of a $Z$-boson, the corresponding EW logs due to the soft/collinear $Z$-boson region need to be consistently removed from our reweighting formulas for the comparison. A successful comparison also relies on a judicious choice of the SM input parameters. The most accurate predictions are obtained using tree-level input parameters in the $G_\mu$-scheme~\cite{Brensing:2007qm,Dittmaier:2009cr}.

Up to now we discussed pure EW corrections, obtained by reweighting tree-level Monte Carlo events. We can straightforwardly combine EW corrections with NLO QCD effects by reweighting the  \PWG~DY generator~\cite{Alioli:2008gx}. The reweighting strategy is similar to the one described in section~\ref{pwg}, with the reweighting factors given by eq.~\eqref{rwgfull}. The only difference is that reweighting now depends also on the $t$ and $u$ Mandelstam variables, and not only on $s$. These are computed on the  \PWG~events, before showering, with the following prescription. If a gluon is present in the final state, we assume that it is emitted from the initial parton moving along the positive $z$-axis if it moves in the right hemisphere (in the center of mass frame), and the converse for left-hemisphere gluons. Concretely, we compute $t$ and $u$ using the four-momentum of the incoming quark or anti-quark that travels in the $z$ direction opposite to the final-state gluon. For gluon-initiated process, the momentum of the initial quark or anti-quark is employed. There is of course no ambiguity in events without emissions. At the leading log in QCD, where the emissions are collinear or soft and factorize, this prescription is exact. It is not exact at NLO, for hard emissions. However the $t$- and $u$-dependent terms in the reweighting are EW corrections, therefore we do not need to model them precisely at NLO in QCD since mixed (two-loops order) QCD and EW corrections are not included in our calculation.

Summarizing, our reweighting produces NLO QCD events, consistently matched with QCD parton showering, and including NLL EW corrections on the SM and on the new physics contributions. The NLL QED accuracy for partially exclusive quantities, like lepton momenta defined with a narrow recombination cone, or ``bare'' muons momenta, is obtained by the {\sc{Pythia}}~8~\cite{Sjostrand:2014zea} QED showering. We validated QED showering effects by comparing with the literature~\cite{Li:2012wna,CarloniCalame:2007cd,Barze:2013fru,Dittmaier:2009cr}. In particular we reproduced table~$1$ of ref.~\cite{Dittmaier:2009cr}, with $\Delta{R}^{\rm{rec}}\hspace{-2pt}=\hspace{-2pt}0.1$ recombination cone, $p_{T,{\rm{min}}}^\gamma\hspace{-2pt}=\hspace{-2pt}10$~GeV, and $|\eta|^\gamma_{\rm{max}}\hspace{-2pt}=\hspace{-3pt}3$ thresholds for photons. The thresholds on the recombined leptons are $p_{T,{\rm{min}}}^l\hspace{-5pt}=\hspace{-2pt}25$~GeV and $|\eta|^l_{\rm{max}}\hspace{-5pt}=\hspace{-3pt}2.5$. We also recombine to the nearest lepton the lepton pairs produced by photon splitting. The same recombination strategy is adopted for the predictions reported in the following sections. 

Before concluding this section, it is worth emphasizing that our result does not include real emissions of massive vector bosons. Namely we target a final state without $W$ or $Z$ bosons. While theoretically well-defined, this final state is not experimentally accessible because the vector bosons might not be detectable if they are soft, or collinear to the beam, or if they decay to neutrinos. We could straightforwardly account for real emissions, including new physics by reweighting, because at the NLL order the real emissions factorize. Therefore they can be generated through splitting, starting from Monte Carlo events without emissions, duly weighted to include new physics. We did not implement this strategy because it is much simpler to use the {\sc{MadGraph}}. For a tree-level process such as the massive vector bosons emission, reweighting is automated and can be used to include the EFT effects. The effect of real corrections depends strongly on the exact definition of the cross-section that is measured experimentally, which in turn is also dictated by experimental considerations. Therefore we ignore real corrections in the analysis of the following section, having in mind an hypothetical measurement of the exclusive cross-section as defined above. However it should be emphasized that these effects should be properly taken into account in the experimental analysis because they are as relevant as the virtual EW logs~\cite{Baur:2006sn}, as expected. Two more processes are not included in our results. One is the photon-quark dilepton production, which does depend on new physics but is extremely small~\cite{Dittmaier:2009cr,Brensing:2007qm}. The other is photon-photon initiated production, which is not sensitive to new physics and thus can be easily added on top by a tree-level SM simulation.

\section{The Drell-Yan Likelihood}\label{DYL}

We now turn to phenomenological applications. In this section we discuss the parametrization of the predicted cross-section as a function of W and Y, with the associated uncertainties, and use it to build the binned Likelihood function needed for the W\&{Y} interpretation of the DY measurements. This will be the starting point for the LHC sensitivity projections reported in section~\ref{sec:pro}.

\subsection{Cross-section parametrization}\label{choleski}

Suppose the neutral and charged DY cross-sections are measured in bins, labeled by the index $I$. The theoretical expected cross-section, denoted by $\sigma_{I}^{\text{th}}$, is a quadratic polynomial in the parameters of interest W and Y. The cross-section is positive, so it can be parametrized as 
\bea\label{sth}
&&\sigma_{I}^{\text{th}}\({\rm{W}},{\rm{Y}}\) =  {\overline\sigma}_{I}^{\text{\sc{sm}}} c_{0,I}^2 \left|
\left(
\begin{array}{ccc}
 1 & c_{1,I} & c_{3,I}\\ 0 & c_{2,I} & c_{4,I}  \\ 0& 0 &c_{5,I}
\end{array}
\right)\cdot
\left(\begin{array}{c}1\\ {\rm{W}}\\ {\rm{Y}}\end{array}\right)
\right|^2\\
&&\qquad\qquad\quad=\overline\sigma_{I}^{\text{\sc{sm}}} c_{0,I}^2 [1+2 c_{1,I} {\rm{W}} +2  c_{3,I} {\rm{Y}}+
(c_{1,I}^{2}+c_{2,I}^{2}){\rm{W}}^2 +\left(c_{3,I}^{2}+c_{4,I}^{2}+c_{5,I}^{2}\right){\rm{Y}}^2 \nonumber\\
&&\qquad\qquad\quad \;\;\;\;\qquad+2( c_{1,I} c_{3,I}+ c_{2,I}c_{4,I}){\rm{W}} {\rm{Y}}  ]\,.\nonumber
\eea
by employing the Cholesky decomposition for positive $3\times3$ matrices, in terms of six dimensionless coefficients $c_{k,I}$, with $k=0,\ldots,5$. The decomposition is unique provided $c_{0,I}$, $c_{2,I}$ and $c_{5,I}$ are positive, while both signs are allowed for the other coefficients. In the equation, $\overline\sigma_{I}^{\text{\sc{sm}}}$ denotes the prediction for the SM cross-section in each bin evaluated with central-value inputs. Namely, with the strong coupling constant $\as$ set to the central value $\as=0.1180$ and with central-value PDF and renormalization/factorization scales. Consequently, the central value of the ${c}_{0,I}$ coefficients is equal to one by definition: $\overline{c}_{0,I}=1$.

The central values of the other coefficients, $\overline{c}_{k,I}$, are readily computed with our reweighted samples, starting from central-value SM Monte Carlo data. As explained in the previous section, the reweighted events contain the coefficients of the weights as a polynomial in W and Y. These are summed up in each bin producing the polynomial coefficients in the bin, out of which the Cholesky decomposition coefficients can be computed, provided the cross-section is a positive polynomial as it must be by consistency. This is always the case in the kinematical regimes accessible at the LHC, because the negative EW logs are still sufficiently small. The only subtlety is associated with the dependence on Y of the charged DY cross-section. Since the latter emerges only through the EW logs, which we expanded at fixed order in our reweighting formulas, no ${\rm{Y}}^2$ term is present and the cross-section polynomial becomes negative at ${\rm{W}}=0$ for very large Y. While such large values of Y are phenomenologically irrelevant, we solved the problem by adding the ${\rm{Y}}^2$ term to the charged DY reweighting for a fully consistent combined expansion in the new physics and in the EW loop parameters.

\subsection{Parametric and theoretical uncertainties}\label{unc}

We now discuss the estimate of the uncertainties on the theoretical predictions for the $c_{k,I}$ coefficients.
These are described statistically, and included in the Likelihood, in terms of nuisance parameters, with an approach that can fit both in a frequentist and in a Bayesian inference framework. From the frequentist point of view the nuisance are related to parameters the $c_{k,I}$ predictions depends on, such as for instance the value of $\as$ or the PDF. The results of auxiliary measurements (e.g., $\as$ or PDF measurements) are incorporated in the Likelihood as multiplicative terms that depend on the nuisance parameters but not on the parameters of interest (i.e., W and Y). From the Bayesian perspective, the nuisance are random variables (and so in turn the $c_{k,I}$'s), and the likelihood of the auxiliary measurements can be interpreted as their statistical distribution. In what follows we adopt the Bayesian language to describe the auxiliary likelihood associated with the nuisance parameters, but we eventually employ it for a frequentist inference on the W and Y parameters. 
%
%

Notice that the discussion above applies only to systematic uncertainties with an underlying statistical origin. The uncertainties from scale variation instead, and more in general all the uncertainties associated with missing higher order corrections in the predictions, do not possess a robust statistical interpretation. As customary we will nevertheless include them as nuisance parameters, but fortunately we will see that they do not play a dominant role in our sensitivity projections.

We now examine the different sources of uncertainties individually, discuss their parametrization in terms of nuisance parameters, and start quantifying their impact. 

\subsubsection*{Uncertainty from Monte Carlo statistic}

No nuisance parameters must be included for Monte Carlo statistical uncertainties, which are completely negligible. More precisely, the uncertainties on the new physics terms are negligible provided the Monte Carlo statistics is sufficient to provide accurate enough (well below $1\%$) predictions of the SM terms. This is because new physics is included by reweighting, hence the relative accuracy on the new physics $c_{k,I}$ parameters is the same one of the SM terms. Since new physics is itself a correction to the SM in the kinematical regime of interest and for the relevant values of the W and Y parameters, the resulting cross-section uncertainty is completely negligible. Accurate SM predictions for unfolded differential cross-section measurements are easy to obtain. If instead the analysis had to be performed on the observed distribution, producing large enough detector simulations might be problematic. However once this is achieved, new physics effects could be included by reweighting with negligible Monte Carlo error. As discussed in the Introduction, it would have been harder to bring the uncertainties on the new physics prediction to a negligible level if employing Monte Carlo predictions that are not obtained by reweighting.

\subsubsection*{Uncertainty from $\boldsymbol{\as}$}
The uncertainty coming from the value of the QCD coupling $\as$ is, by construction, determined by a single parameter. It is thus included through a single nuisance parameter $\theta^{\as}$ affecting all bins in a correlated way. The nuisance is distributed as a standard normal, i.e.
\beq\l{sigmadistralphaS2}
f_{\as}\(\theta^{\as}\)= \frac{1}{\sqrt{2\pi}}e^{-\f12{\(\theta^{\as}\)^{2}}}\,.
\eeq
We can regard $\theta^{\as}$ as a variable related to the physical $\as$ (which is Gaussian-distributed by assumption) by a suitable linear transformation that brings its distribution to the standard normal. 

\begin{figure}[t]
\centering
\begin{minipage}{0.52\textwidth}
  \centering
  \includegraphics[width=1\linewidth]{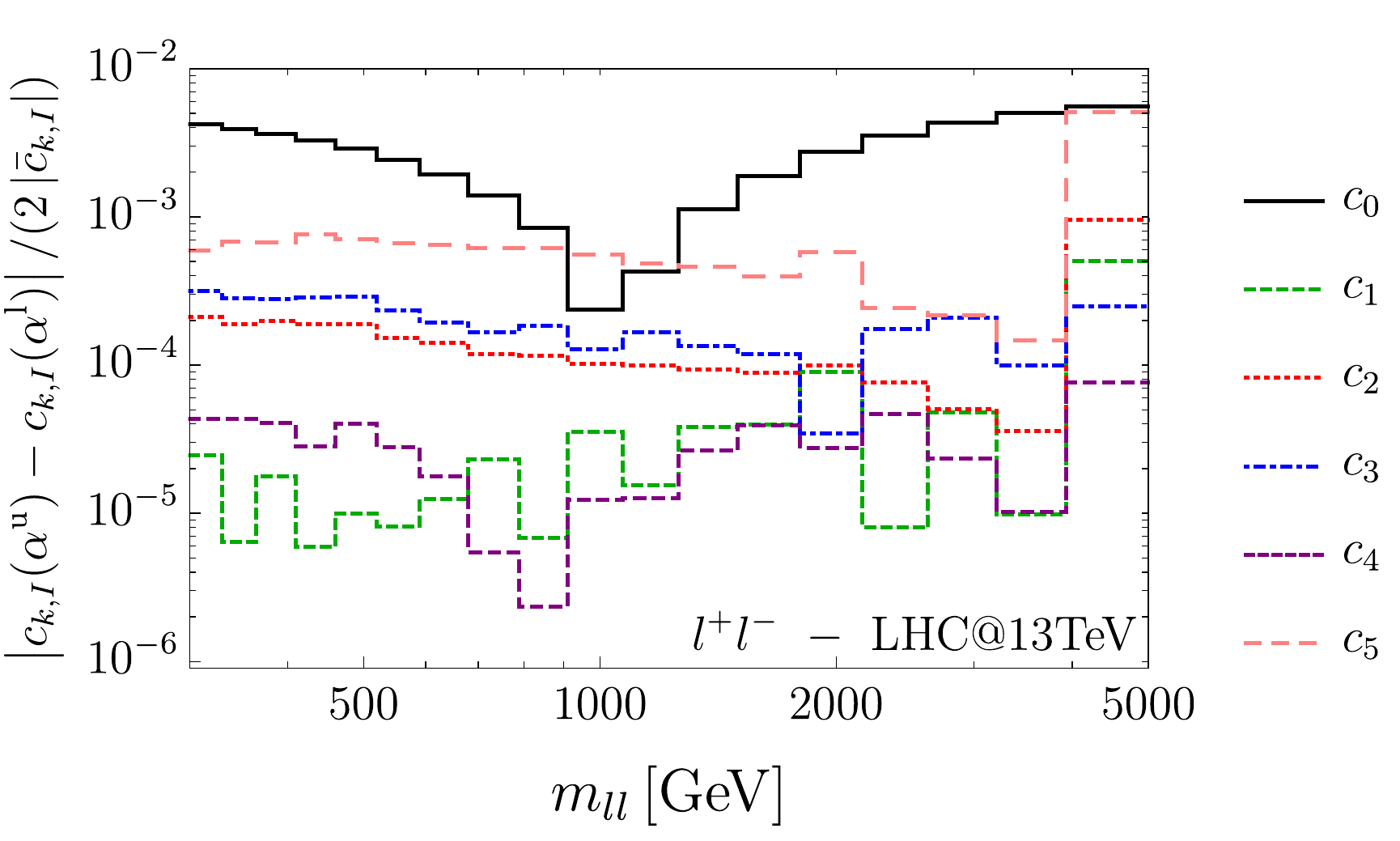}
\end{minipage}\hspace{1mm}
\begin{minipage}{0.465\textwidth}
  \centering
  \includegraphics[width=1\linewidth]{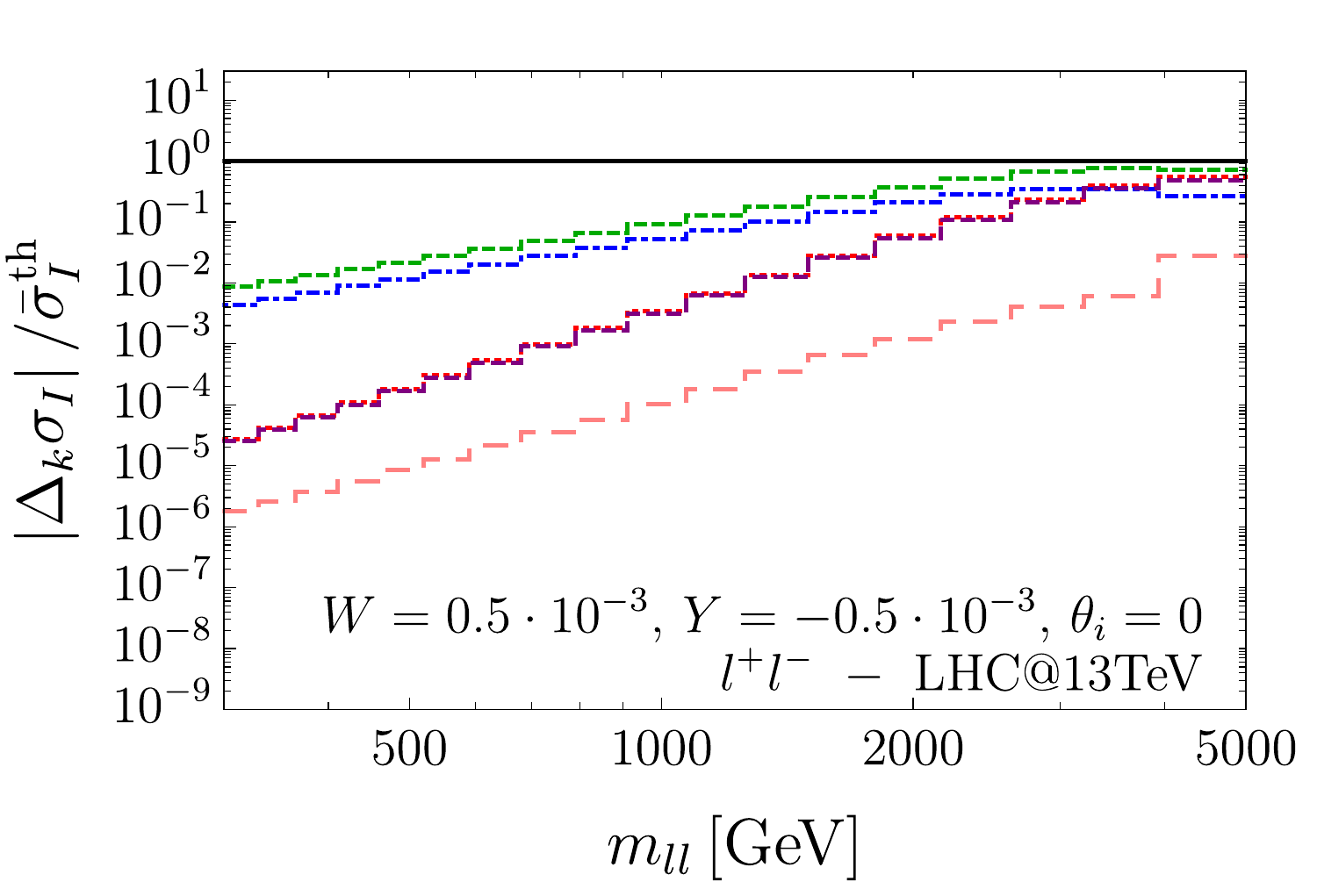}
\end{minipage}
\caption{{Left:} Uncertainties on the $c_{k,I}$ coefficients from variation of the value of $\as$, computed as $\left|c_{k,I}(\as^{\text{u}})-c_{k,I}(\as^{\text{l}})\right|/(2\left|\overline{c}_{k,I}\right|)$. {Right:} Relative impact of the $c_{k,I}$ coefficients on the total $\sigma_{I}^{\text{th}}$ for $W\hspace{-2pt}=\hspace{-2pt}5\cdot 10^{-4}$, $Y\hspace{-3pt}=\hspace{-2pt}-\hspace{-1pt}5\cdot10^{-4}$, and all nuisance parameters $\theta_{i}=0$. This is computed as $|\Delta_k\sigma_I|/\overline\sigma_{I}^{\text{th}}$, with $\Delta_k\sigma_I$ the difference between the central-value cross-section $\overline\sigma_{I}^{\text{th}}$ and the value of $\sigma_{I}^{\text{th}}$ obtained by setting ``$c_k$'' to zero in eq.~\eqref{sth}.} \label{fig:alphaS_unc}
\end{figure}

The \PWG~SM DY~\cite{Alioli:2008gx} Monte Carlo samples include the weights of each event when $\as$ is set to the lower and upper ($\as^{\text{l}}=0.1165$ and $\as^{\text{u}}=0.1195$) boundaries of the $1\sigma$ confidence interval, plus of course the weight for $\as$ equal to its central value $\as=0.1180$. From the latter, we obtain the central-value coefficients ${\overline{c}}_{k,I}$ (with ${\overline{c}}_{0,I}=1$ as previously discussed). From the former, we obtain the values of $c_{k,I}$ for $\as=\as^{\text{l}}$ and for $\as=\as^{\text{u}}$. The resulting relative variations are shown in the left panel of figure~\ref{fig:alphaS_unc} for the neutral DY invariant mass distribution, with the binning employed for the LHC projections in section~\ref{sec:pro}.

We see that the $\as$ uncertainties are rather small, compared with the expected experimental (statistical and systematic) uncertainties of the cross-section measurements (see figure~\ref{BSMwithU}). Also notice that the $\as$ uncertainties are much smaller for the new physics ${{c}}_{k,I}$'s (for $k=1,\ldots,5$) than for the overall multiplicative ${{c}}_{0,I}$ coefficient, which encapsulate in particular the uncertainty on the SM term of the prediction. Moreover, the new physics contribution to the cross-section is small, suggesting that all the $\as$ uncertainties apart from those on ${{c}}_{0,I}$ can safely be ignored in the analysis. This is confirmed by the right panel of figure~\ref{fig:alphaS_unc}, which quantifies the relative impact of the new physics terms to the total expected cross-sections $\sigma_{I}^{\text{th}}$. Large values of the W and Y parameters are chosen in the figure, well above the projected LHC sensitivity with only $100$~fb$^{-1}$. Even for these values, new physics is a small correction to the SM up to around $2$~TeV energies. At this high energy, $\as$ uncertainties are anyhow irrelevant because of the large statistical uncertainties (see again figure~\ref{BSMwithU}). Similar conclusions are reached by studying the charged DY transverse mass distribution we consider in section~\ref{sec:pro} for the LHC projections.

In light of the above discussion, we include the dependence on $\theta^{\as}$ only on ${{c}}_{0,I}$, with a linear parameterization
\beq\l{sigmaSM_alphaS}
c_{0,I}=c_{0,I}\(\theta^{\as}\)={\overline{c}}_{0,I}+\kappa_{I}^{\as}\theta^{\as}=1+\kappa_{I}^{\as}\theta^{\as}\,,
\eeq
where the $\kappa_{I}^{\as}$'s are computed as
\beq
k_{I}^{\as}=\max\(\left| c_{0,I}(\as^{\rm{u}}) - {\overline{c}}_{0,I}\right|, \big| c_{0,I}(\as^{\rm{l}}) - {\overline{c}}_{0,I}\big|\)\,.
\eeq
If the dependence of the coefficients on $\as$ was exactly linear, the upper and lower variations would be exactly equal and opposite, and eq.~\eqref{sigmaSM_alphaS} would describe exactly the dependence of $c_{0,I}$ on $\as$. We have verified that the variations are equal and opposite to good approximation, and the maximal variation was selected for conservative results. Notice that the parameterization in eq.~\eqref{sigmaSM_alphaS} does not respect the condition $c_{0,I}>0$ for the unicity of the Cholesky decomposition. This does not produce negative cross-sections, but (formally) results in a double coverage of the space of the predictions in terms of $\theta^{\as}$. However the problem is irrelevant in practice because the uncertainties are so small that $c_{0,I}$ will never change sign in the Likelihood marginalization (or profiling) process.

\subsubsection*{Uncertainty from the parton distribution functions}

The PDF uncertainties on the $c$'s are computed using \PWG, with the same strategy outlined above for the $\as$ uncertainties. We employed the $30$ PDF in the set \textsc{PDF4LHC15\_nlo\_30\_pdfas} (code 90400 in the LHAPDF database~\cite{Buckley:2014ana}) \cite{Butterworth:2015oua,Dulat:2015mca,Harland-Lang:2014zoa,Carrazza:2015hva}, which correspond to the Hessian reduction of the PDF uncertainties to $30$ nuisance parameters $\theta_{i}^{\text{\sc{pdf}}}$, with $i=1,\ldots,30$. The nuisance are uncorrelated and normally distributed:
\beq\l{nuis_pdf_1}
f_{\text{\sc{pdf}}}\(\theta_{i}^{\text{\sc{pdf}}}\)= \frac{1}{\sqrt{2\pi}}e^{-\f12{\(\theta_{i}^{\text{\sc{pdf}}}\)^{2}}}\,,\;\;\;\;\;i=1,\ldots,30\,.
\eeq
The use of an Hessian set is legitimated by the fact that we look for small deviations from the SM, rather than to on-shell new physics. In this context, the Hessian parametrization allows for a simpler treatment of the PDF uncertainties including correlations between different bins and different process such as the neutral and charged DY. Our choice of the set with $30$ replicas, in alternative to the one with $100$ replicas, is motivated by a study we performed for neutral DY using the \textsc{PDF4LHC15\_nlo\_mc\_pdfas} Monte Carlo ensemble set, where we identified less than $20$ eigenvectors of the $c$'s covariance matrix with uncertainties above $\permil$.

The following dependence of the $c_{k,I}$ coefficients on the PDF nuisance parameters is assumed. The $c_{0,I}$, $c_{2,I}$ and $c_{5,I}$, which need to be positive for the unicity of the Cholesky decomposition, are parametrized with an exponential:
\beq\label{pdf_parametrisationexp}
X(\theta_{i}^{\text{\sc{pdf}}})={\overline{X}}\exp\left[\sum\limits_{i=1}^{30}
\frac{X^{(i)}-{\overline{X}}}{{\overline{X}}}
\theta_{i}^{\text{\sc{pdf}}}\right]\,,\;\;{\rm{for}}\;\;X=\{c_{0,I},c_{2,I},c_{5,I}\}\,.
\eeq
The others are parameterized linearly
\beq\label{pdf_parametrisationlin}
X(\theta_{i}^{\text{\sc{pdf}}})={\overline{X}}+ \sum\limits_{i=1}^{30} (X^{(i)}-{\overline{X}}) \theta_{i}^{\text{\sc{pdf}}}\,,\;\;{\rm{for}}\;\;X=\{c_{1,I},c_{3,I},c_{5,I}\}\,.
\eeq
In the equation, we indicate with a bar the central value predictions, while the superscript $^{(i)}$ denotes the value of the parameter obtained with each of the $30$ PDF replicas in the set. The parametrization is such that $X$  equals (approximately, in the case of eq.~\eqref{pdf_parametrisationexp}) $X^{(i)}$ when $\theta_{i}^{\text{\sc{pdf}}}$ is at its one-sigma value and all the other $\theta_{i}^{\text{\sc{pdf}}}$'s vanish, compatibly with the definition of the Hessian set.

The PDF uncertainties are larger than those on $\as$, and eventually turn out to be the dominant component of the total theoretical uncertainties shown in figure~\ref{BSMwithU}. Furthermore these uncertainties grow with the energy like the new physics effects. Therefore in our analysis we account for them fully, both in the SM and in the new physics contributions to the cross-section.

\subsubsection*{Uncertainty from missing higher orders}\label{miss}

The uncertainties due to the truncation of the perturbative series in the cross-section prediction are harder to quantify, and impossible to incorporate rigorously in any statistical framework. Nevertheless we can estimate their impact as follows. Missing higher orders in the QCD loop expansion are estimated by varying the factorization ($\mu_F$) and QCD-coupling renormalization ($\mu_R$) scales independently around the central values $\overline\mu_F=\overline\mu_R=\sqrt{s}$. The scales are varied by multiplicative factors equal to $2^{\pm1}$, ${2}^{\pm1/2}$ and $1$, in a grid with a total of $24$ entries plus the central value configuration. The maximal and the minimal values of the $c_{k,I}$ coefficients in this grid, denoted as $c_{k,I}^{\rm{max}}$ and $c_{k,I}^{\rm{min}}$ below, are used for the uncertainty estimate. Missing higher order in the EW loop expansion are instead estimated by adding the leading IR logarithmic terms at two loops to our reweighting formulas. All IR logs have been computed in ref.~\cite{Denner:2006jr} at two loops, however only the leading (i.e., $L^4$ angular-independent) terms are retained in the estimate of the uncertainties. Compatibly with Sudakov resummation formulas,  these are straightforwardly included by replacing $f_{\rm{a.i.}}\rightarrow f_{\rm{a.i.}} + f_{\rm{a.i.}}^2/2$ in eq.~\eqref{FD}. The predictions for the $c_{k,I}$ coefficients that include this contribution are denoted as $c_{k,I}^{2-\text{Sudakov}}$.

\begin{figure}[t]
\centering
\begin{minipage}{0.52\textwidth}
  \centering
  \includegraphics[width=1\linewidth]{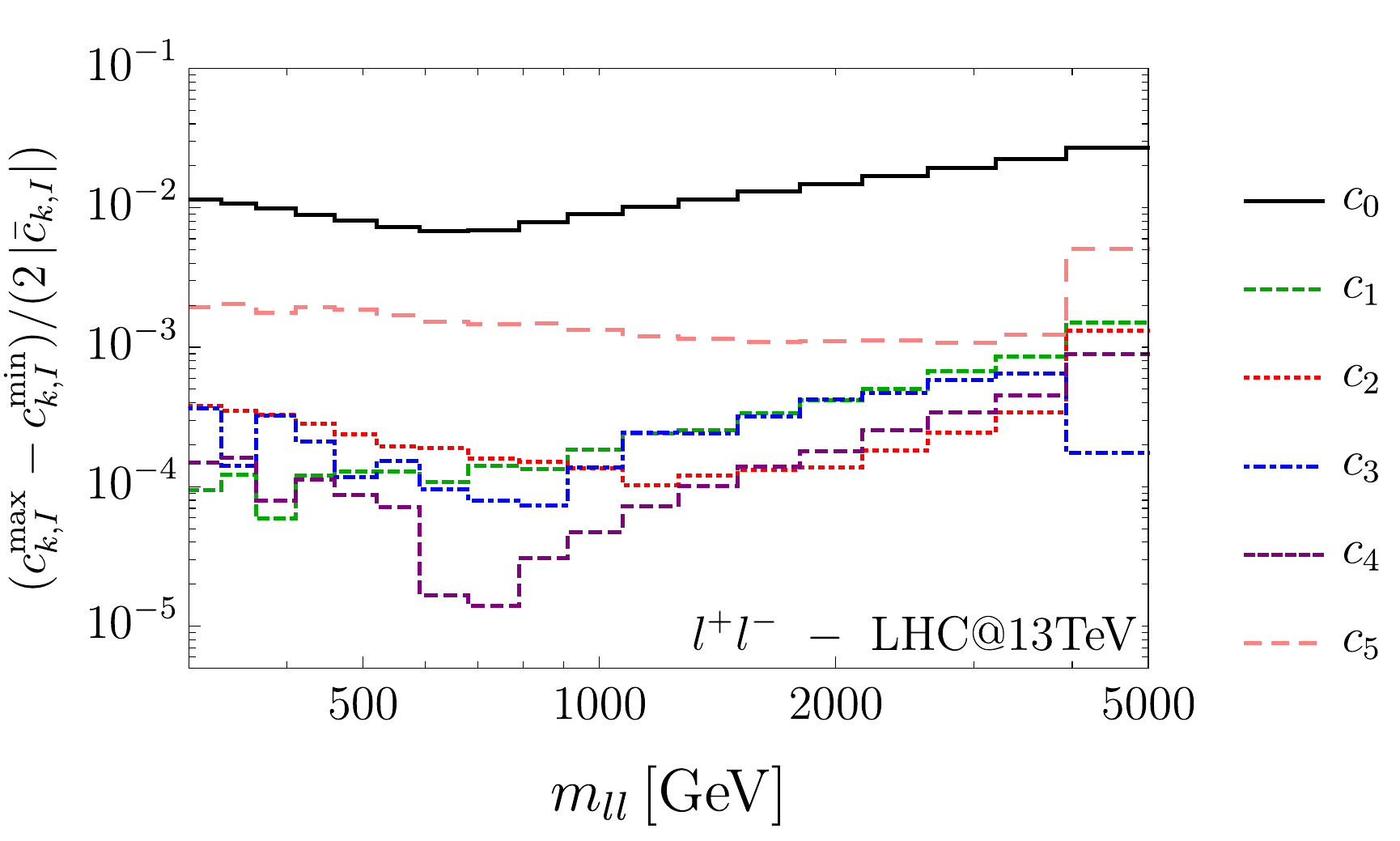}
\end{minipage}\hspace{1mm}
\begin{minipage}{0.465\textwidth}
  \centering
  \includegraphics[width=1\linewidth]{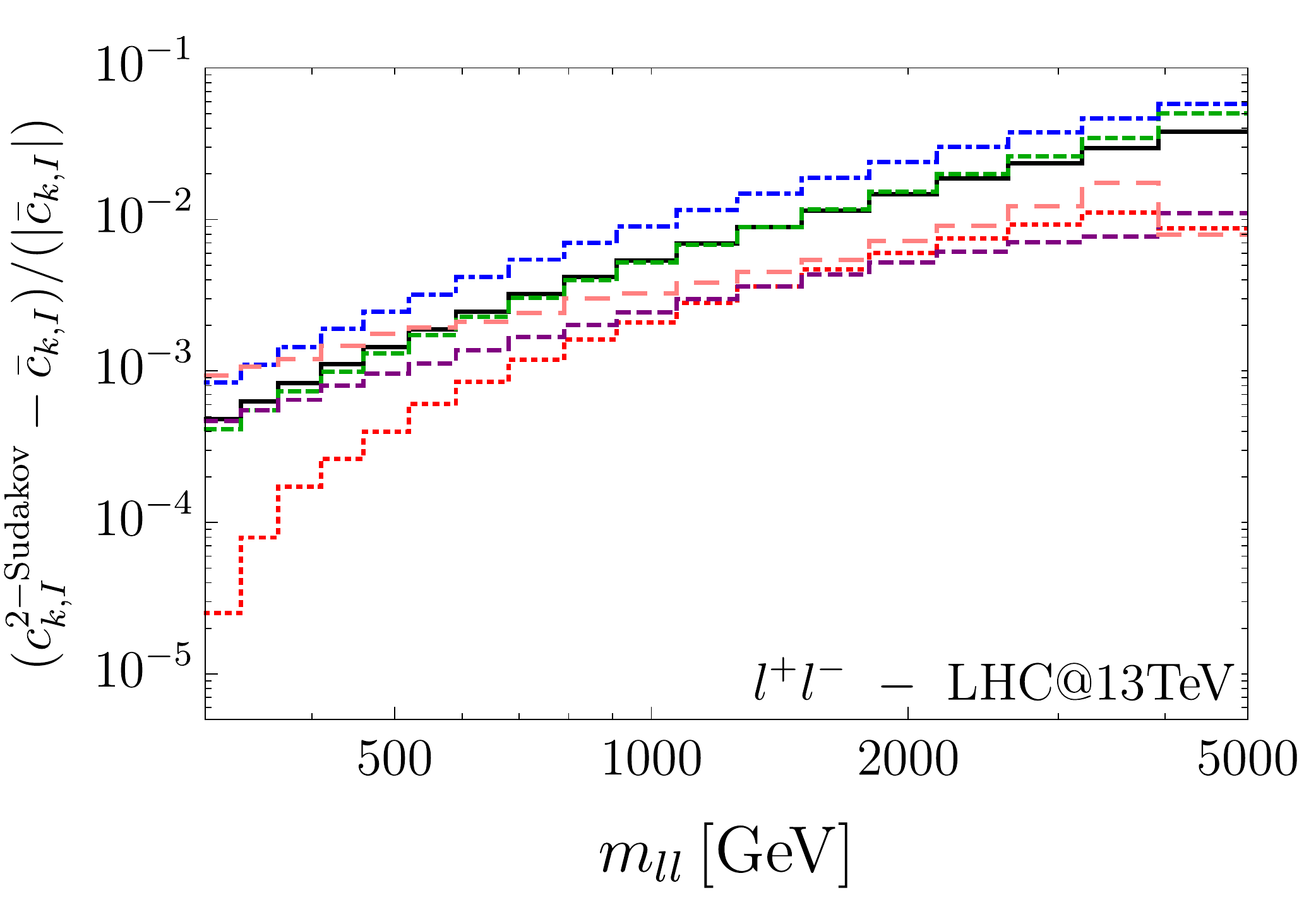}
\end{minipage}
\caption{{Left:} Uncertainties from scale variation computed as $(c_{k,I}^{\rm{max}}-c_{k,I}^{\rm{min}})/(2{\overline{c}}_{k,I})$. {Right:} Uncertainties from missing EW loops estimated as explained in the main text.} \label{fig:SV_unc}
\end{figure}

The uncertainties from missing higher orders in QCD (left panel) and in EW (right panel) are displayed in figure~\ref{fig:SV_unc}. We discuss them in turn. NLO QCD scale variation effects are known (see, e.g., ref.~\cite{Alioli:2016fum}) to be sizable in the SM. Correspondingly we see in the figure that the uncertainties on the $c_{0,I}$'s are relatively big. On the other hand, the scale uncertainties on the new physics $c_{k,I}$'s (with $k\neq0$) are extremely small and completely negligible. Namely, we find that the NLO QCD scale variations mostly affect $\sigma_{I}^{\text{th}}$ in eq.~\eqref{sth} as an overall new physics-independent multiplicative factor. The uncertainties due to the missing higher-orders in the EW loop-expansion are smaller than the QCD scale variation, and they become sizable only at high energy where the statistical error gets big. They are definitely irrelevant for the new physics term, but they could play a role for the SM contribution, in particular for the charged DY process where they are slightly larger than what shown in the figure for the neutral case.

The previous results show that our predictions for the new physics contribution to the cross-section are sufficiently accurate, and the associated theoretical uncertainties can be neglected. On the SM term instead, NLO QCD scale variations and missing higher orders in the EW expansion are potentially relevant. However the SM predictions are available at NNLO~\cite{Li:2012wna}, and $2$-loops enhanced EW logarithms can be easily included in by analytic reweighting. By replacing the SM term of our prediction with the NNLO SM, and including $2$-loops logs, we could thus lower the NLO scale variations to the NNLO level, and make higher order EW corrections completely negligible. In what follows we will thus ignore EW effects and include NNLO-sized QCD scale variations which we estimate, following ref.~\cite{Li:2012wna}, to be one tenth of the NLO ones. These uncertainties are modeled  by introducing one nuisance parameter $\theta_I^{\text{\sc{tu}}}$ for each bin, following a standard normal distribution. Linear dependence on $\theta_I^{\text{\sc{tu}}}$ is assumed for $c_{0,I}$
\beq\label{th_parametrisation}
{{c}}_{0,I}={\overline{c}}_{0,I}+\frac{(c_{k,I}^{\rm{max}}-c_{k,I}^{\rm{min}})}2
\theta_I^{\text{\sc{tu}}}\,.
\eeq

\subsection{Statistical inference}


In the following section we will present sensitivity estimates for the W\&{Y} parameters at the LHC with the standard~\cite{Cowan:2010js} frequentist approach based on the profile Likelihood ratio and employing Asymptotic formulas and the ``Asimov dataset''. Namely, we define the ``$t_\mu$'' test statistic (with $\mu=({\rm{W}},{\rm{Y}})$ the parameters of interest), with the Likelihood in the numerator maximized over the nuisance parameters for fixed W and Y and the one in the denominator maximized also on the parameters of interest. The Asymptotic ($\chi^2_2$) distribution is assumed for $t_\mu$ in the EFT hypothesis in order to set the $95\%$ (or $68\%$)~CL boundaries, while the median $t_\mu$ in the SM hypothesis is obtained by setting the observed data to the central-value SM prediction. 

The treatment of experimental (statistical and systematical) uncertainties would be completely straightforward if the experimental result was presented as a measurement of the unfolded cross-section in the bins. Namely, the complete Likelihood will be merely obtained by plugging $\sigma_{I}^{\text{th}}$ in the experimental Likelihood, expressed as a function of the ``truth-level'' cross-sections $\sigma_I$, including the dependence on the parameters of interest and on the nuisance, and multiplying by the nuisance parameters constraint terms. The simplest way to mimic the complete Likelihood would be to employ a Gaussian guess for the experimental Likelihood, which should include an estimate of the uncertainties on the measurement emerging from the combination of statistical and systematic errors. Since it is unclear how the statistical and systematic errors should be combined, a slightly more sophisticated approach is considered in what follows. However it should be emphasized that this adds nothing to the accuracy of our modeling of the experimental errors, given the lack of basic information on the systematic uncertainties expected in the measurement and of the (potentially very important) correlations between the errors in different bins and in neutral and charged DY. One advantage of the strategy we follow is that it could be adapted to the direct comparison of the W and Y prediction with the observed-level distributions without unfolding. 

\subsubsection*{Experimental uncertainties}\l{exp_syst}

The experimental Likelihood for the $\sigma_I$ cross-sections emerges from the number of events, $n_I$, observed in each bin. These are Poisson-distributed independent variables with means $\mu_I$ that are related to the theoretical predictions $\mu_{I}^{\text{th}}={\text{L}}\cdot\sigma_{I}^{\text{th}}$ (with ${\text{L}}$ the integrated luminosity) up to experimental uncertainties which we encapsulate in normal-distributed nuisance parameters $\theta_I^{\text{exp}}$, and to the luminosity uncertainty. Namely, the $\mu_I$ are defined as
\beq\label{mui}
\mu_I=\mu_{I}^{\text{th}}\left(1+\sum_J\left[\sqrt{\Sigma^{\text{exp}}}\right]_{I}^{\;\;J}\theta_J^{\text{exp}}+{0.02}\,{\theta^{\text{L}}}\right)\,,
\eeq
where $\Sigma^{\text{exp}}$ is the covariance matrix associated with the systematic experimental uncertainties in the relation between the truth-level expected countings $\mu_{I}^{\text{th}}$ and the observed-level expectations $\mu_I$. Notice that the expression above does not take into account event migrations from the truth- and observed-level bins, which should be encapsulated in the response matrix that multiplies the $\mu_{I}^{\text{th}}$ term. However it can model realistically the effect of uncertainties on the determination of the response matrix, provided a reasonable guess is made for the covariance matrix $\Sigma^{\text{exp}}$. The simple choice we consider in the next session is based on current experimental results. The error on the luminosity measurement, at the $2\%$ level, is described by the normally distributed nuisance parameter $\theta^{\text{L}}$.

The complete Likelihood we will employ for the statistical inference finally reads
\beq\label{likefull}
\bry{lll}
\dst \mathscr{L}\(W,Y,\theta^{\as},\theta_{i}^{\text{\sc{pdf}}},\theta_{I}^{\text{\sc{tu}}},\theta_I^{\text{exp}},\theta^{\text{L}}\) &=&\dst \prod_{I=1}^{N}\text{Poisson}\left[n_{I}| \mu_{I}\(W,Y,\theta^{\as},\theta_{i}^{\text{\sc{pdf}}},\theta_{I}^{\text{\sc{tu}}},\theta_I^{\text{exp}},\theta^{\text{L}}\)\right]\vspace{2mm}\\
&& \times f_{\as}(\theta^{\as}) f_{\text{\sc{pdf}}}(\theta^{\text{\sc{pdf}}}) f_{\text{\sc{tu}}}(\theta_{I}^{\text{\sc{tu}}})f_{\text{exp}}(\theta^{\text{exp}})f_{\text{L}}(\theta^{\text{L}})\,.
\ery
\eeq
The dependence of $\mu_I$ (through $\sigma_I^{\rm{th}}$, as in eq.~\eqref{mui}) on the parametric and theoretical uncertainties is introduced by combining additively the correction terms in eqs.~\eqref{sigmaSM_alphaS}, \eqref{pdf_parametrisationexp}, \eqref{pdf_parametrisationlin}, \mbox{and \eqref{th_parametrisation}.}

\begin{figure}[t]
\centering
\begin{minipage}{0.5\textwidth}
  \centering
  \includegraphics[width=1.1\linewidth]{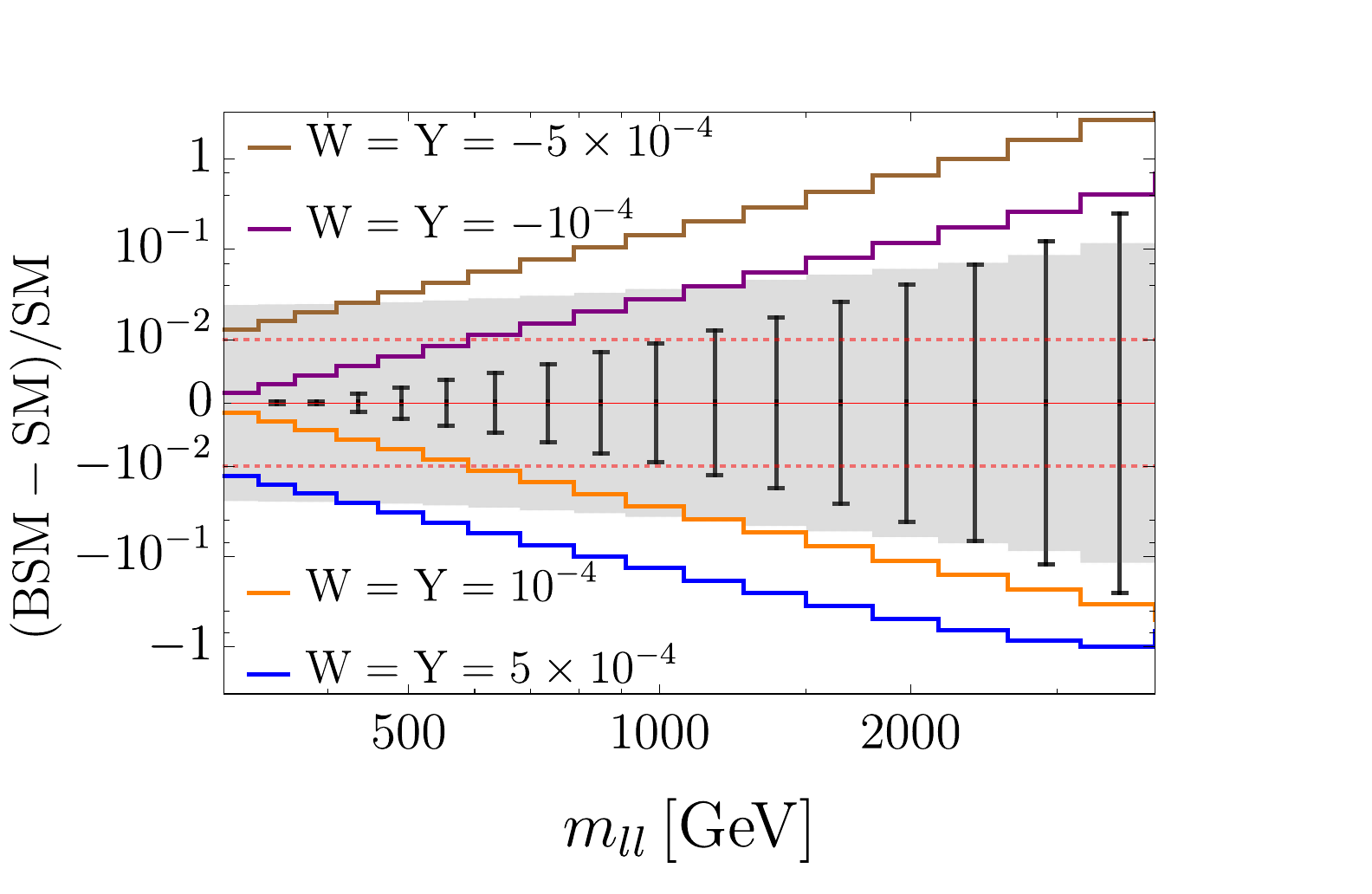}
\end{minipage}\hfill
\begin{minipage}{0.5\textwidth}
  \centering
  \includegraphics[width=1.1\linewidth]{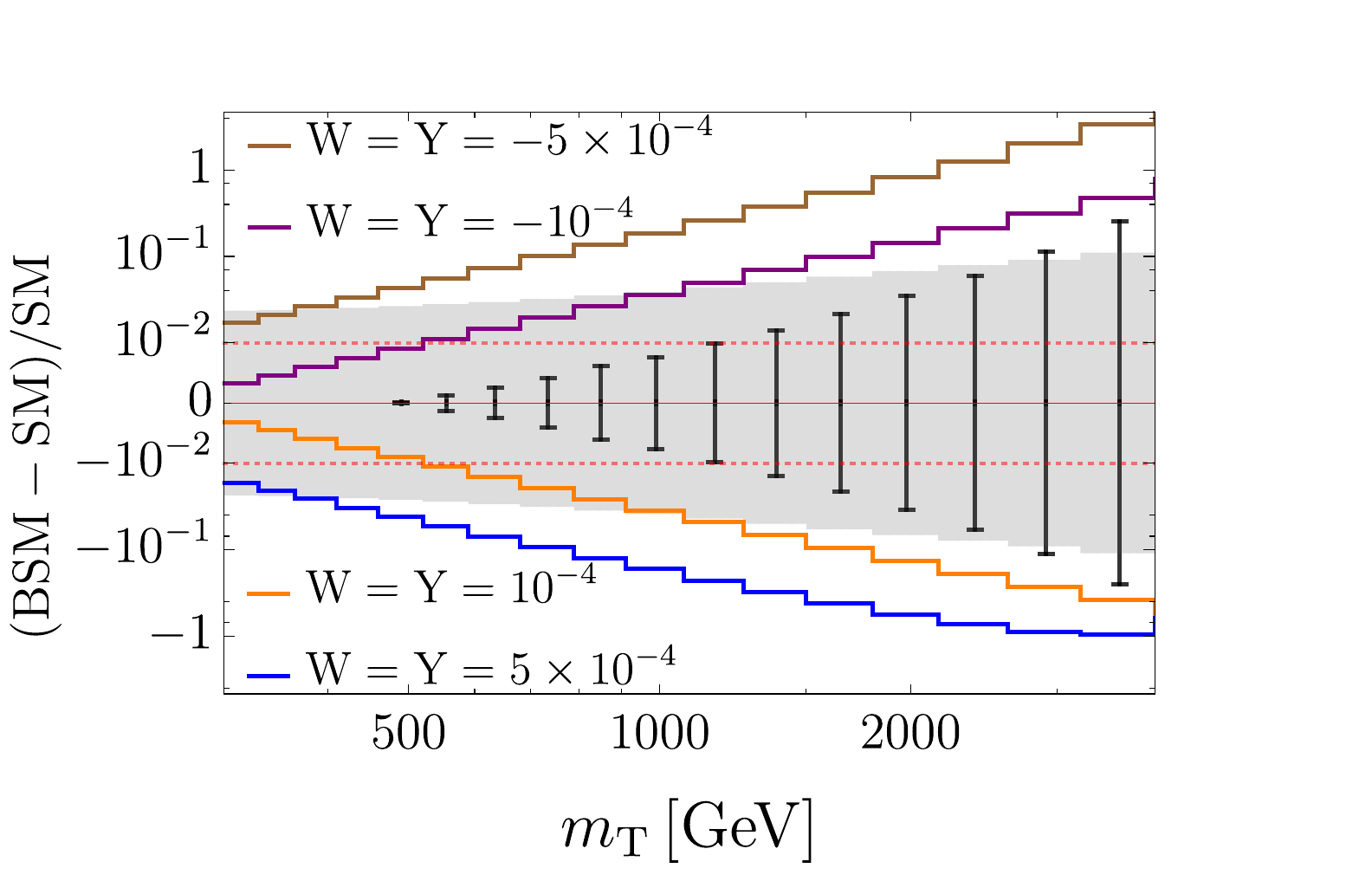}
\end{minipage}
\caption{Invariant and transverse mass distribution of the relative discrepancy between BSM and SM predictions, for neutral (left panel) and charged (right panel) Drell-Yan. The gray band represents the uncertainties (at $1\sigma$) in the theoretical predictions, while the black bars denote the statistical uncertainties estimated as one over the square root of the number of expected events. The HL-LHC integrated luminosity (${\rm{L}}= 3\,{\rm{ab}}^{-1}$) is assumed. \label{BSMwithU}}
\end{figure}

\section{LHC projections}\label{sec:pro}
We base our projection on hypothetical measurements of the neutral DY invariant mass ($m_{ll}$) and of the charged DY transverse mass ($m_{T}$) distributions in logarithmically-spaced bins
\beq\nonumber
\left\{.3,.33,.365,.41,.46,.52,.59,.68,.79,.91,1.07,1.26,1.5,1.8,2.16,2.62,3.20,3.93,13\right\}{\rm{TeV}}\,.
\eeq
The LHC collider energy is set to $13$~TeV, and integrated luminosities of $100\,{\rm{fb}}^{-1}$, $300\,{\rm{fb}}^{-1}$, and $3\,{\rm{ab}}^{-1}$ are considered. The former luminosity is roughly the one that has been collected as of today. The two latter ones are those that will be available at the end of the LHC and of the HL-LHC programs, respectively. We incorporate in the projections $65\%$ identification efficiency for electrons and $80\%$ for muons, which effectively reduces the luminosity by a factor of $2$ in neutral DY and by around $40\%$ for the charged process.

The projections are obtained with the Likelihood described in the previous section, where all the relevant sources of parametrical and theoretical uncertainties in the cross-section predictions are taken into account. However they are not fully realistic because the experimental systematic uncertainties in the cross-section measurements (and the correlation of these uncertainties across different bins) can only be estimated by the experimental collaboration. Based on run-$1$ results, in our ``baseline'' scenario we set to $2\%$ and to $5\%$ the experimental relative uncertainties in the measurement of the neutral and of the charged cross-sections, respectively. No correlation is assumed across different bins, i.e. $\Sigma^{\text{exp}}\propto \mathds{1}$ in eq.~\eqref{mui}, aiming to a conservative result.

The results are illustrated in the rest of this section, starting from those in the baseline configuration for the uncertainties. We next consider departures from the baseline setup and discuss the impact of the various sources of uncertainties separately.

\begin{figure}[t]
\centering
\begin{minipage}{0.315\textwidth}
  \centering
  \includegraphics[height=1\linewidth]{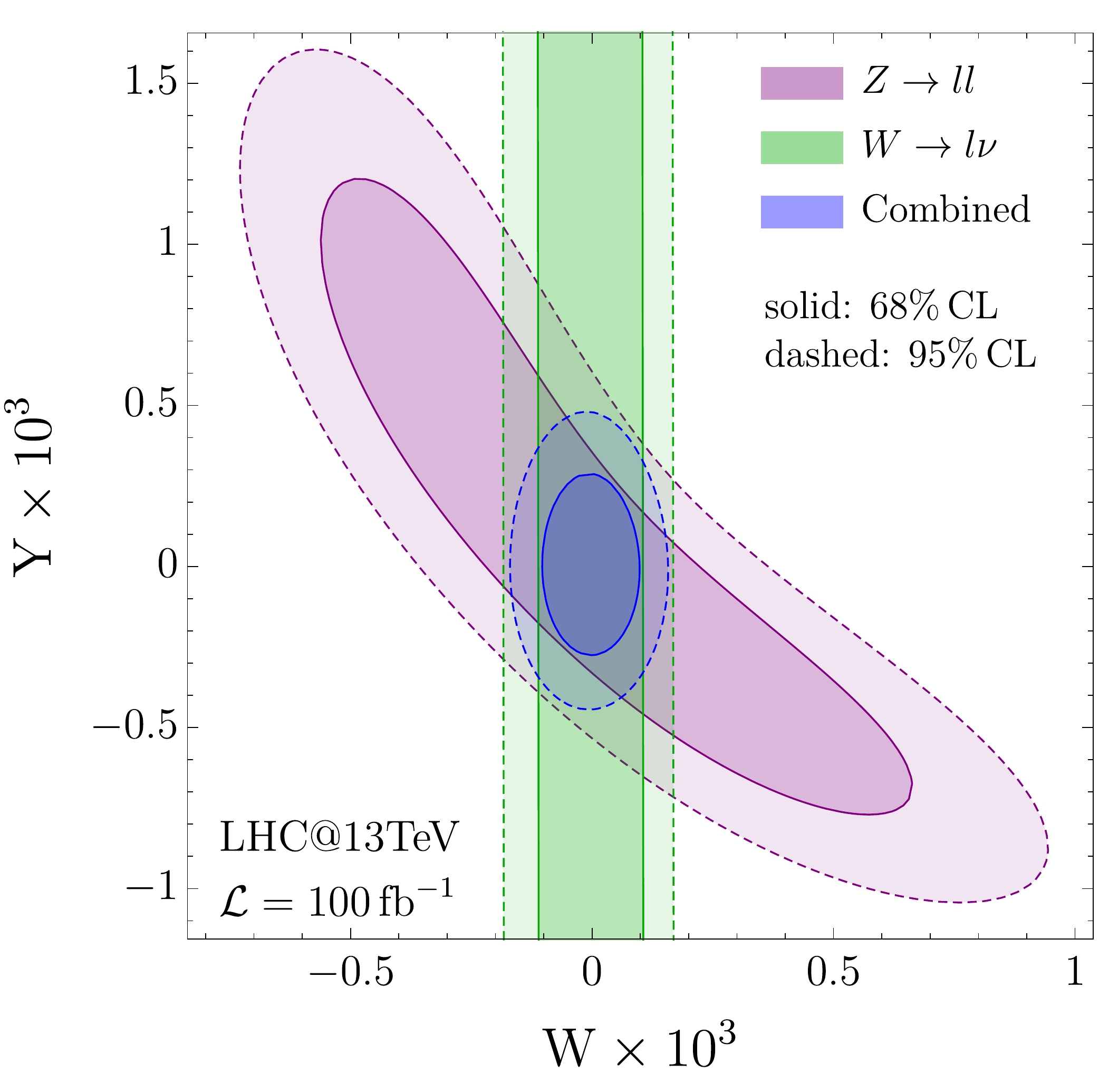}
\end{minipage}\hfill
\begin{minipage}{0.315\textwidth}
  \centering
  \includegraphics[height=1\linewidth]{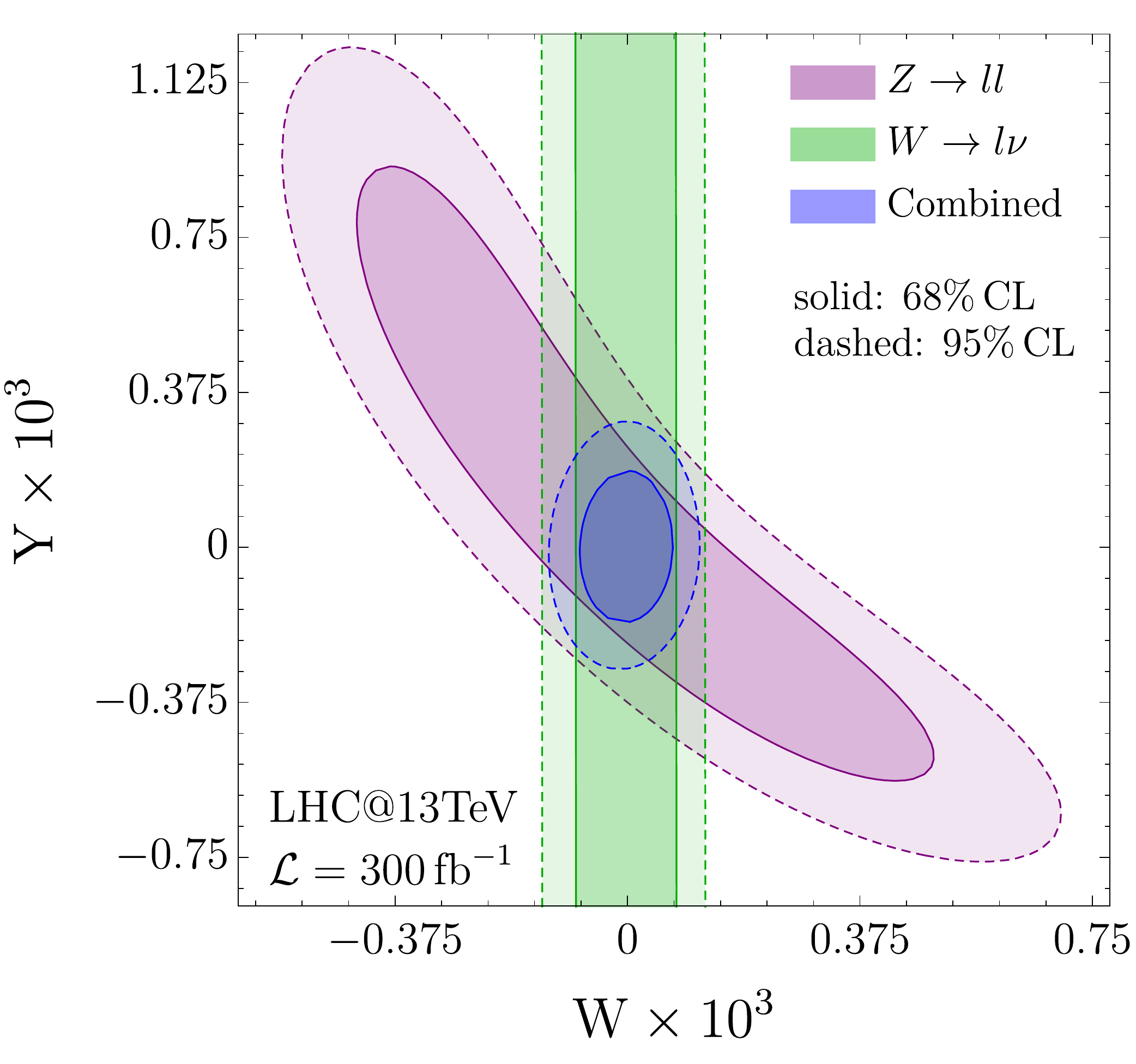}
\end{minipage}\hfill
\begin{minipage}{0.315\textwidth}
  \centering
  \includegraphics[height=1\linewidth]{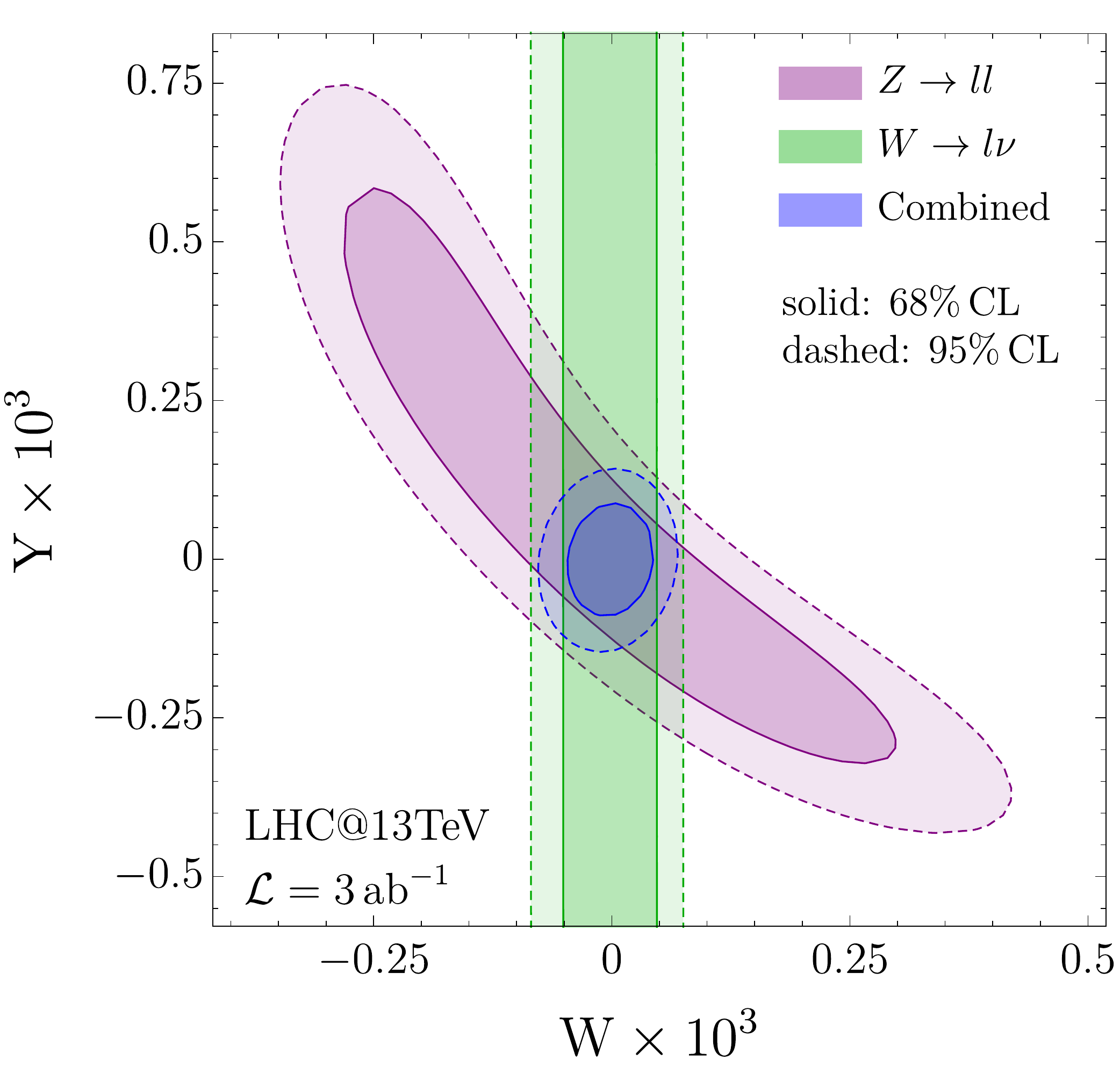}
\end{minipage}%
\caption{Projected $68\%$ and $95\%$ exclusions in the W\&{Y} plane for different luminosities from neutral (purple) and charged (green) Drell-Yan measurements at $13$~TeV LHC.} \label{fig:2DFreq}
\end{figure}

A first qualitative assessment of the sensitivity can be obtained by looking at figure~\ref{BSMwithU}. The figure shows the corrections to the cross-section, relative to the SM, at $4$ points in the W\&{Y} parameter space, overlaid with the total uncertainties in the theoretical predictions, represented as a gray shaded region. As discussed in the previous section, these uncertainties are dominated by the PDF contribution. The black bars correspond to the statistical uncertainties in each bin at the HL-LHC. The $1\%$ uncertainty level is marked with horizontal dotted lines because it provides a reasonable absolute lower bound to the systematic component of the experimental error on the cross-section measurements, on top of the statistical one. Based on the figure, we expect  values of W\&{Y} of the order of $1\cdot10^{-4}$ or less to be within the reach of the HL-LHC. This is confirmed by the contours in the W\&{Y} plane, at $68$ and $95\%$~CL, displayed in figure~\ref{fig:2DFreq} for the $3$ integrated luminosities we considered. The neutral and charged DY sensitivities are shown separately and combined.

\begin{figure}[h!]
\centering
\begin{minipage}{0.485\textwidth}
  \centering
  \includegraphics[width=1\linewidth]{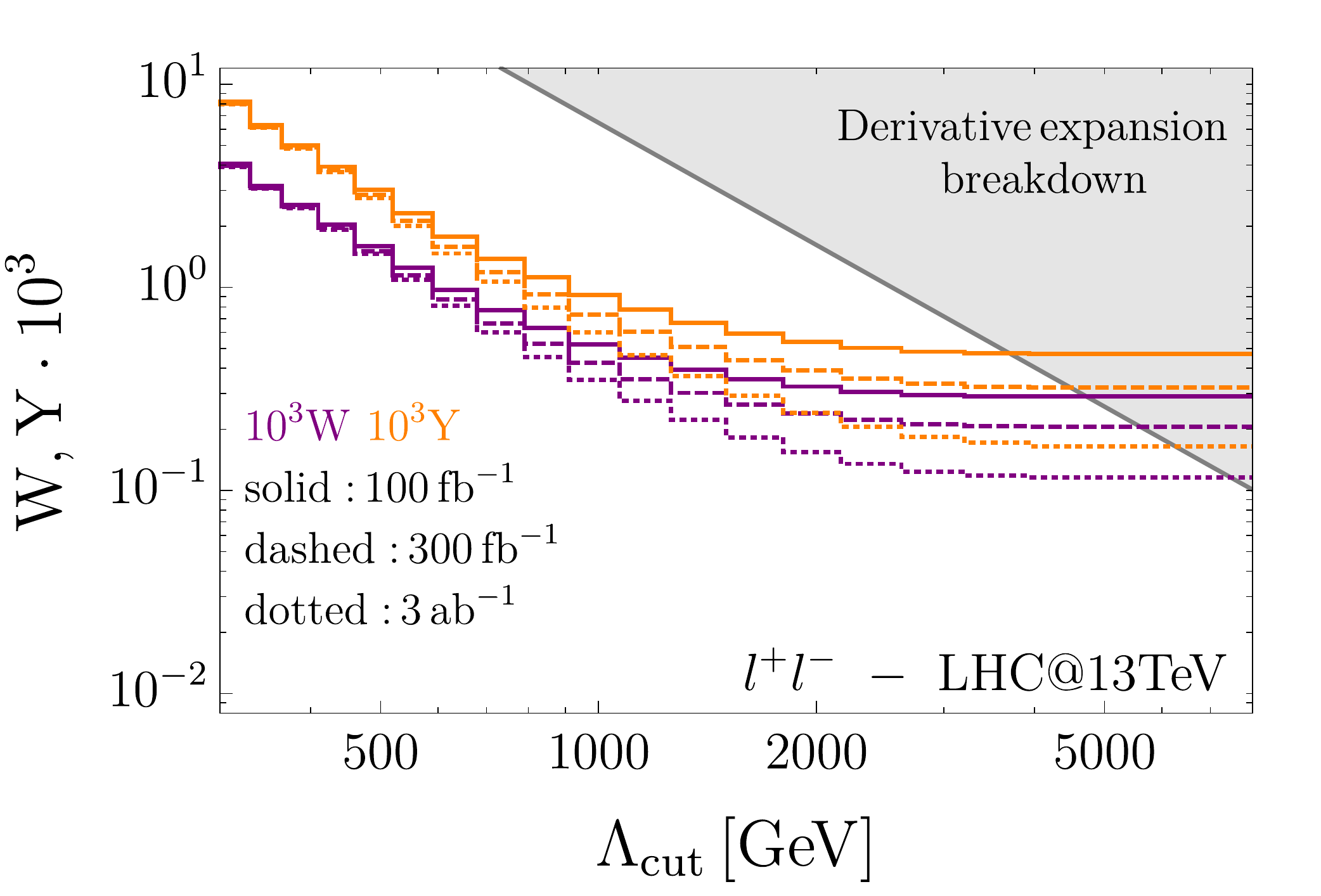}
\end{minipage}\hfill
\begin{minipage}{0.485\textwidth}
  \centering
  \includegraphics[width=1\linewidth]{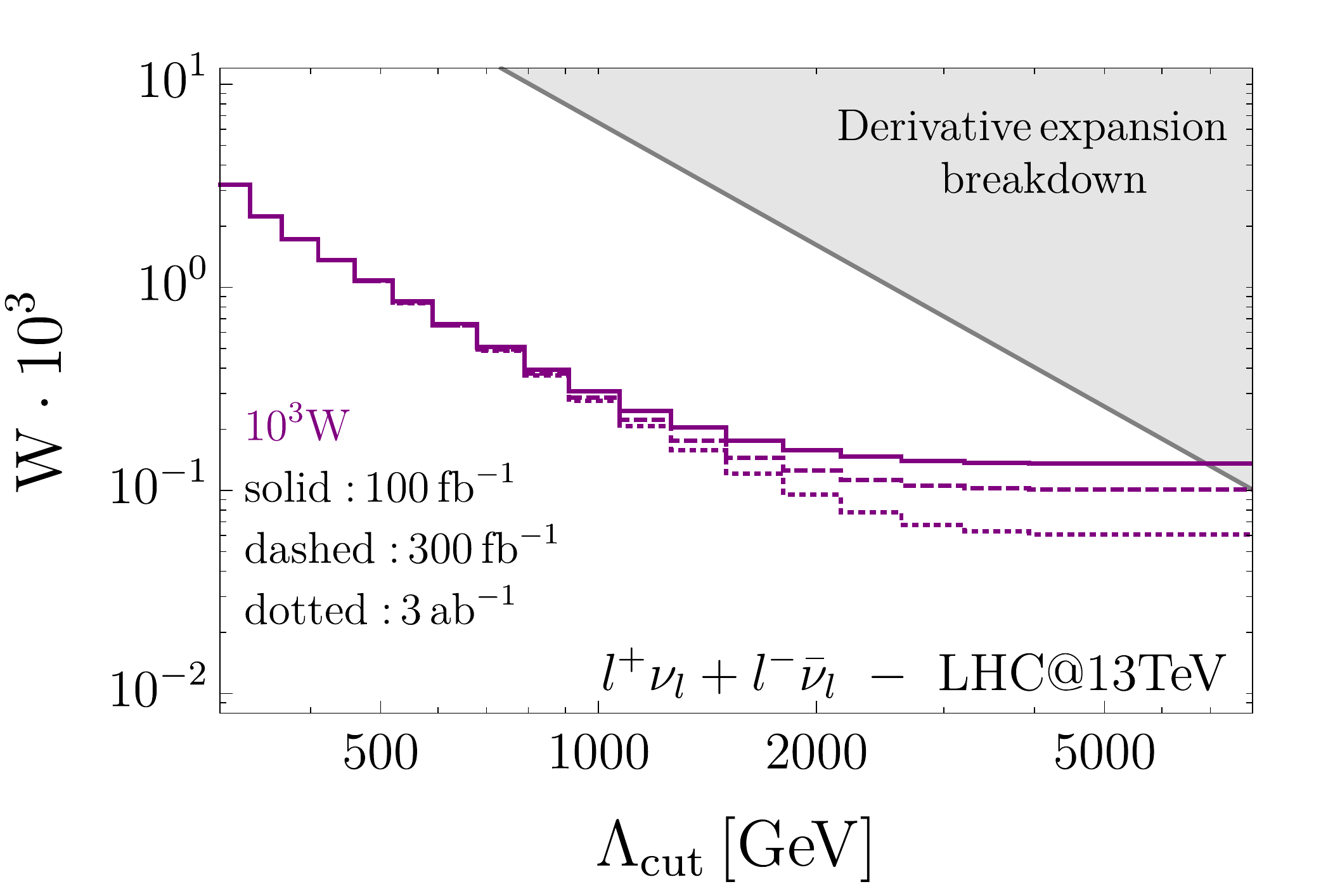}
\end{minipage}
\begin{minipage}{0.485\textwidth}
  \centering
  \includegraphics[width=1\linewidth]{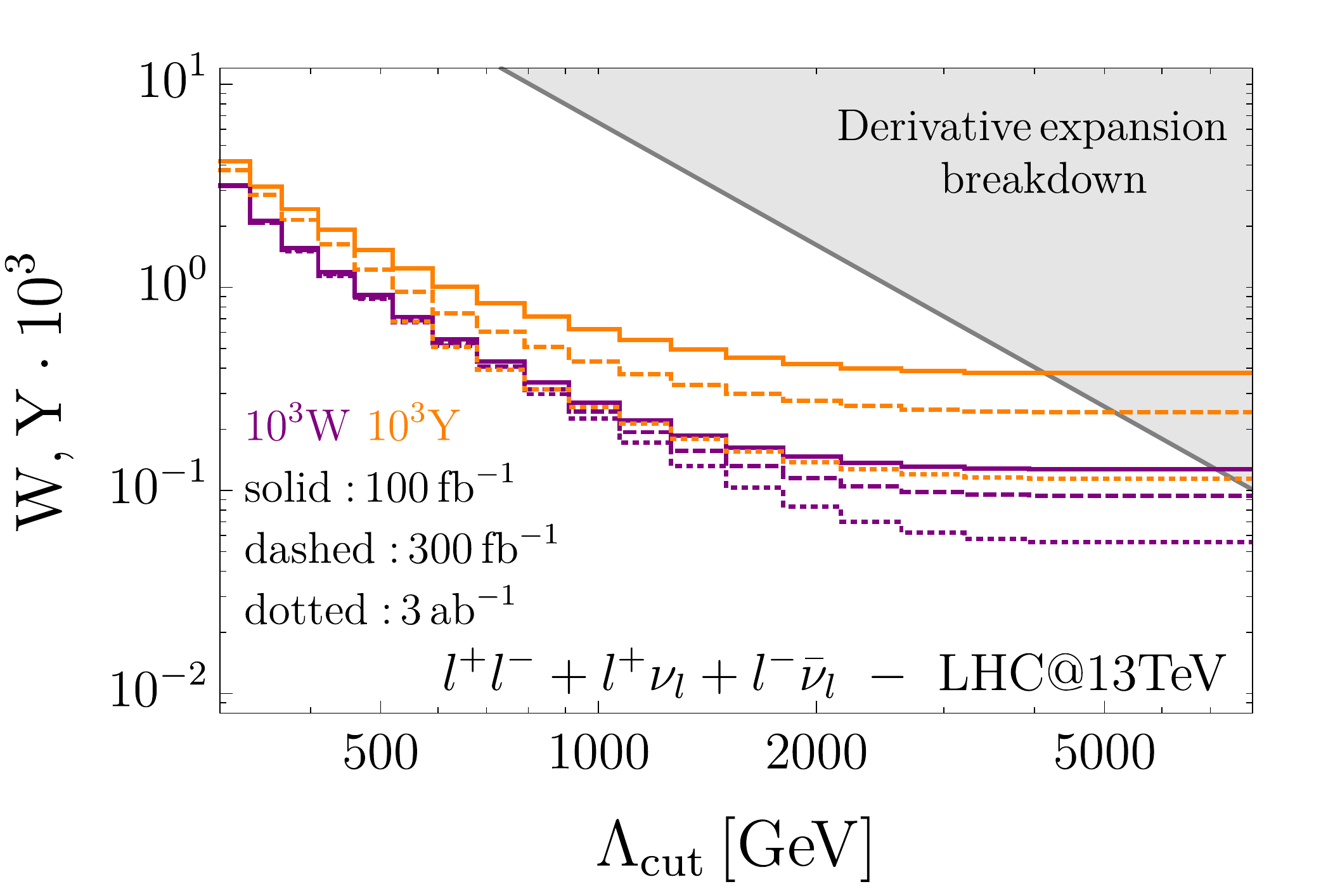}
\end{minipage}
\begin{minipage}{0.485\textwidth}
  \centering
  \includegraphics[width=1\linewidth]{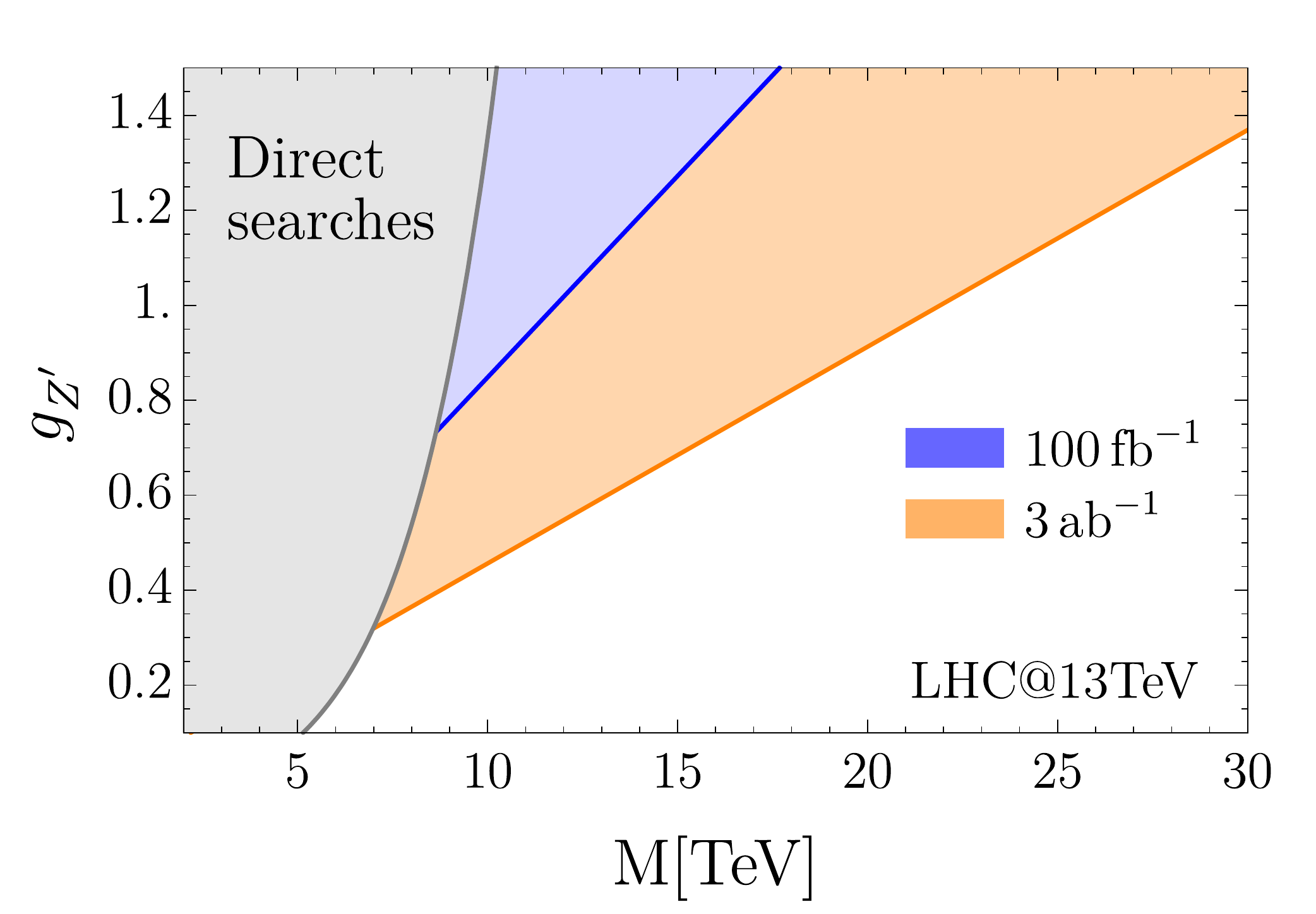}
\end{minipage}
\caption{Projected  bounds  as  a  function  of  a  cutoff  on  the  mass  variable. Bottom right:~Projected exclusions on a simple $Z^\prime$ model (defined as in ref.~\cite{Strategy:2019vxc}) from the measurement of the Y parameter. The exclusion reach from direct $Z^\prime$ searches, at the HL-LHC, is also shown. \label{fig:ladder}}
\end{figure}

We further inspect our results, following ref.~\cite{Farina:2016rws}, from the viewpoint of the validity of the EFT modeling of new physics. The first three plots in figure~\ref{fig:ladder} show single-parameter $95\%$~CL sensitivities as a function of the maximal energy (invariant or transverse mass) of the data employed in the analysis. These are obtained considering only one (W or Y) parameter of interest, with the other set to zero. The first two panels refer to neutral and charged DY, respectively, and the third one to the combination of the two channels. Consistently with ref.~\cite{Farina:2016rws}, we see that the reach sits comfortably below the ``Derivative Expansion Breakdown" region, showing that the usage of the EFT is justified and the resulting limits are valid. More quantitatively, we see that the energy region which is relevant for the limit does not exceed $2$ or $3$~TeV. The value of the W\&{Y} parameters we are sensitive to can easily be due to new physics particles which are much heavier than that. For instance in Composite Higgs theories the new physics scale could easily be at $10$~TeV or more, justifying the usage of the EFT at few TeV energies. A more simple examples is the one of a $Z^\prime$, such as the ``Universal $Z^\prime$ model'' employed in ref.~\cite{Strategy:2019vxc} for future colliders performance assessments. The sensitivity projection on the Y parameter (which is the only one generated by this model), once translated in the mass-coupling plane of the $Z^\prime$ model as in figure~\ref{fig:ladder}, reveals that the HL-LHC could be sensitive to values of the Y parameter induced by a $Z^\prime$ which is as heavy as $30$~TeV. The projected direct reach on the $Z^\prime$ particle at the HL-LHC, from ref.~\cite{Strategy:2019vxc}, is overlaid to the figure in order to outline that the sensitivity to the model is dominated by the neutral DY measurement of Y in a wide region of the parameter space.

\begin{figure}[h]
\centering
  \centering
  \includegraphics[width=0.95\linewidth]{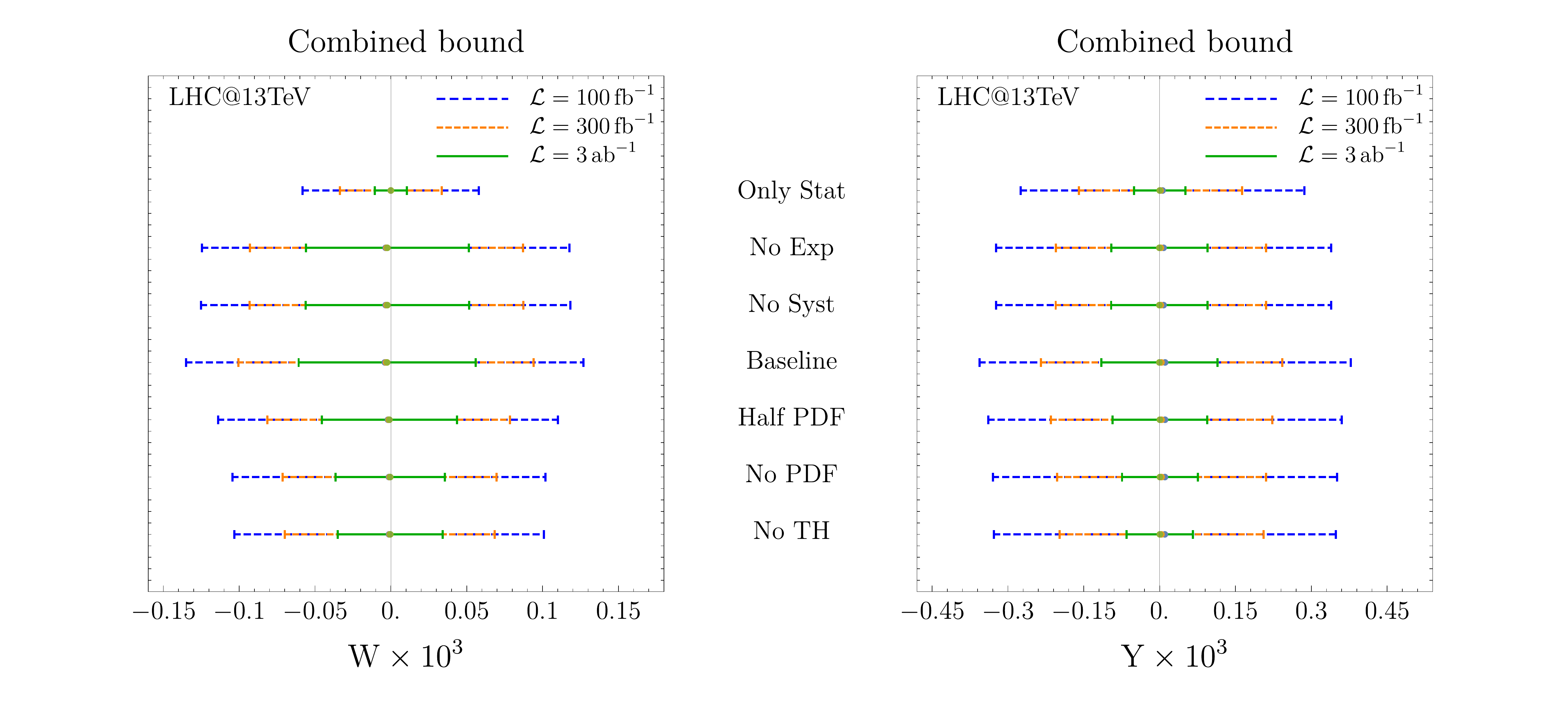}
\caption{Single-parameter $95\%$ reach on W (left) and on Y (right), with different integrated luminosities and for different uncertainty configurations.
\label{Uncplot}}
\end{figure}

It is interesting to investigate the impact of each source of uncertainty on the sensitivity. We report in figure~\ref{Uncplot} the projected single-operator limits obtained with different assumptions on the errors, compared with the baseline configuration. Eliminating the uncertainties from $\as$ and from missing higher orders in the perturbative expansion is found not to improve the sensitivity appreciably, and for this reason the corresponding reach is not reported in the figure. On the other hand, we have verified that the reach would significantly deteriorate with respect to the baseline, especially at the HL-LHC, if the theory uncertainties on the SM prediction were increased to the level estimated by the NLO scale variation in figure~\ref{fig:SV_unc}. Incorporating uncertainties from missing $2$-loops EW Sudakov effects degrades instead the reach by $10\%$ at most at the HL-LHC. The baseline reach projections thus rely on the availability of NNLO predictions, while it is less relevant to include the enhanced EW logarithms at the $2$-loops order.

As expected, PDF are the most relevant source of uncertainties in the theoretical predictions. However we see in figure~\ref{Uncplot} that halving or eliminating these uncertainties does not improve the sensitivity radically. We now turn to the uncertainties of experimental origin, i.e. the luminosity and the $\Sigma^{\text{exp}}$ uncertainty of eq.~(\ref{mui}). Removing the latter (as in the ``No Syst'' bars) has a moderate impact on the reach, while the former is completely irrelevant. Indeed, by removing also the luminosity uncertainty (``No Exp'' bar), the reach does not improve further. 

The picture emerging from the previous discussion is that the experimental accuracy assumed in the baseline configuration is sufficient, given the state-of-the art accuracy of the theoretical predictions, to exploit at best the LHC and HL-LHC potential to probe the W\&{Y} parameters, and vice versa. A more accurate determination of the PDF could improve the sensitivity, but only slightly. On the other hand, it should be emphasized that our estimate of the experimental uncertainties is a mere guess, which in particular does not take into account correlations between the experimental errors in the different bins and in the different processes, which might reduce the impact of these uncertainties on the reach. If this was the case, the adequacy of the theoretical predictions should be reconsidered, and an improvement of the PDF determination could entail a much more significant progress in the sensitivity. The absolute lower bound for the reach is provided by the ``Only Stat'' bars in figure~\ref{Uncplot}, where all sources of theoretical and experimental systematic error are eliminated.

Before concluding, we compare our results with the findings of ref.~\cite{Farina:2016rws}. Our projected limits are weaker by around $30\%$, due to a different estimate of the PDF uncertainties. In the present paper, we used LHAPDF, which combines several PDF sets, while the estimate in ref.~\cite{Farina:2016rws} was based on ref.~\cite{Alves:2014cda}, where only one set (NNPDF) was considered. The PDF uncertainties of ref.~\cite{Alves:2014cda} are a factor of around $2$ smaller than ours in the relevant kinematical range, making the uncertainties employed in ref.~\cite{Farina:2016rws} effectively correspond to our ``Half PDF'' configuration. With this configuration we could indeed accurately reproduce the results of ref.~\cite{Farina:2016rws}.

\section{Conclusions}\label{sec:con}

We have shown that the effect of the most relevant dimension-$6$ operators (i.e., those that grow quadratically with the energy at the interference level) can be incorporated in the high-energy Drell--Yan predictions by analytic reweighting, up to the NLO accuracy in QCD and including double and single log-enhanced EW corrections at one loop. Our method allows to compute the dependence on the new physics parameters of the cross-section in any phase-space bin without performing a scan on the parameters space. It can also generate events that include QCD and QED showering effects consistently, based on the \PWG~method. 

Two operators in this set, associated with the W and Y parameters, are particularly interesting because they are generated in universal new physics scenarios including Composite Higgs. We thus focused on these operators for an illustration of the methodology, and performed LHC (and HL-LHC) sensitivity projections. Our results confirm and strengthen the findings of ref.~\cite{Farina:2016rws}, where less accurate predictions and systematic uncertainties estimates were employed. The accuracy of our predictions for the new physics contribution to the cross-sections is found to be totally adequate, and the associated uncertainties are negligible. The relevant uncertainties are those on the SM term, and PDF are the dominant source. Theoretical uncertainties are under control provided NNLO QCD predictions are employed for the SM term. One-loop EW radiative corrections should also be included, possibly exactly rather than at the single-log order using our strategy. The impact of two-loops EW logarithms on the reach has been found to be marginal, also at the HL-LHC. Nevertheless, these terms could be included straightforwardly by analytic reweighting.

Our work could be extended in two directions. First, by including all the relevant operators in view of a global EFT fit. Second, by assessing the impact on the sensitivity of the angular distributions, to be studied in multi-differential cross-section measurements. Our analytic reweighting strategy will be crucial for these extensions as it allows to deal with a larger number of parameters and of bins with limited extra computational effort.

\section*{Acknowledgments:}
We thank J.~Ruderman, G.~Panico, D.~Pappadopulo and M.~Farina for useful discussions and for collaboration at the initial stage of this work. We are indebted with A.~Vicini, E.~Mergetti and especially with S.~Alioli for their support on \PWG. We also thank E.~Rizvi, J.~Alcaraz, G.~Cuomo, A.~Glioti and A.~Manenti for discussions. The work of L.R. was supported by the Swiss National Science Foundation under contract 200021-178999. This work was partly supported by the Italian Ministry of Research (MIUR) under the
PRIN grants 20172LNEEZ and 2017FMJFMW.


\appendix
\section{Reweighting factors}\label{RW}

The first term in eq.~\eqref{rwlog} takes the form
\beq
\rho_{{\rm{n}}({\rm{c}}),\Lambda}^{q_1{\overline{q}}_2\rightarrow l_1 {\overline{l}}_2}= \left(1+a_{W,\Lambda}^{\rm{n}(\rm{c})}W+a_{Y,\Lambda}^{\rm{n}(\rm{c})}Y\right)^2\,,\nonumber\\
\eeq
where
\begin{align*}
 a_{W,\Lambda}^{\rm{n}} & = a_W^{\rm{n}}(s;q_{\chi_q},\,l_{\chi_l}) + \frac{\log (\Lambda^2/s)}{16 \pi^2 m_W^2} \frac{ g^2 \beta_{n,W}(q_{\chi_q}, l_{\chi_l}) }{{\mathcal{C}}^0_{\rm{SM}}(q_{\chi_q},\,l_{\chi_l}) } \,, \\
 a_{Y,\Lambda}^{\rm{n}} &= a_Y^{\rm{n}}(s;q_{\chi_q},\,l_{\chi_l}) + \frac{\log (\Lambda^2/s)}{16 \pi^2 m_W^2} \frac{g'^2 \beta_{n,Y}(q_{\chi_q}, l_{\chi_l})}{{\mathcal{C}}^0_{\rm{SM}}(q_{\chi_q},\,l_{\chi_l}) } \,, \\
 a_{W,\Lambda}^{\rm{c}} & = a_W^{\rm{c}}(s)+ \frac{\log (\Lambda^2/s)}{16 \pi^2 m_W^2}\frac{V_{ud}^* g^2 \beta_{c,W}}{{\mathcal{C}}^+_{\rm{SM}}(s;\{{\rm{u}},{\rm{d}}\},\,l)) } \,, \\
  a_{Y,\Lambda}^{\rm{c}}& =  \frac{\log (\Lambda^2/s)}{16 \pi^2 m_W^2}\frac{V_{ud}^* g^{'2} \beta_{c,Y}}{{\mathcal{C}}^+_{\rm{SM}}(s;\{{\rm{u}},{\rm{d}}\},\,l)) }  \,,
\end{align*}
with $a_{W(Y)}^{{\rm{n}}({\rm{c}})}$ and the ${{\mathcal{C}}}$'s defined as in section~\ref{fo}. The $\beta$-functions, computed by {\tt{DsixTools}}~\cite{Celis:2017hod}, are reported in table~\ref{beta}.

\begin{table}[h]
\begin{center}
\renewcommand{\arraystretch}{2}
\begin{tabular}{|c|c|c|c|c|c|c|c|}
\hline
$(q_{\chi_q},l_{\chi_l})$ & $(u_L,e_L)$ & $(d_L,e_L)$ & $(q_L,e_R)$ & $(u_R,e_L)$ & $(d_R,e_L)$ & $(u_R,e_R)$ & $(d_R,e_R)$
\\
\hline
$\beta_{n,W}$ & $\frac{{g^\prime}^2-g^2}{12}$&$\frac{5 \left(11 g^2-{g^\prime}^2\right)}{12}$&$-\frac{1}{6}{g^\prime}^2$&$-\frac{1}{3}{g^\prime}^2$&$\frac{1}{6}{g^\prime}^2$&$0$&$0$ \\
\hline
$\beta_{n,Y}$ & $ \frac{51 g^2-703 {g^\prime}^2}{324}$&$\frac{ -51 g^2-703 {g^\prime}^2}{324}$&$ -\frac{883}{162} {g^\prime}^2$&$ -\frac{853}{81}{g^\prime}^2$&$ \frac{679}{162}{g^\prime}^2$&$ -\frac{1256}{81}{g^\prime}^2$&$ \frac{940}{81}{g^\prime}^2$ \\
\hline
\multicolumn{2}{|c|}{$ \beta_{c,W} = \frac{28 g^2 - 3 {g^\prime}^2}{6}$}&\multicolumn{2}{c|}{$\beta_{c,Y}= -\frac{17}{54}g^2$}&\multicolumn{4}{c}{}
\\
\cline{1-4}
\end{tabular}
\end{center}
\caption{The relevant $\beta$-functions.
\label{beta}}
\end{table}

The second term in eq.~\eqref{rwlog} is given by
\begin{align*}
&\begin{aligned}
\Delta\rho^{u{\overline{u}}\rightarrow l {\overline{l}}}_{\rm{NLL},\Lambda} = 2 \Bigg( \rho_{{\rm{n}},\Lambda}^{u_{\chi_u} \bar{u}_{\chi_u} \rightarrow l_{\chi_l} {\overline{l}}_{\chi_l}} \mathcal{F}_D + \frac{{\delta \mathcal{C}}^0_{\rm{SM}}(u_{\chi_u},\,l_{\chi_l}) }{{ \mathcal{C}}^0_{\rm{SM}}(u_{\chi_u},\,l_{\chi_l}) }\sqrt{\rho_{{\rm{n}},\Lambda}^{u_{\chi_u} \bar{u}_{\chi_u} \rightarrow l_{\chi_l} {\overline{l}}_{\chi_l}}}  \, +
\\
\delta_{\chi_u,L}\delta_{\chi_l,L}\sqrt{\rho_{{\rm{n}},\Lambda}^{u_{\chi_u} \bar{u}_{\chi_u} \rightarrow l_{\chi_l} {\overline{l}}_{\chi_l}} \rho_{{\rm{c}},\Lambda}^{\phantom{A}}}\frac{g^2}{(4 \pi)^2} L_u \Re\left[ \frac{
V_{{\rm{u}}{\rm{d}}^\prime} {\mathcal{C}}^+_{\rm{SM}}(s;\{{\rm{u}},{\rm{d}^\prime}\},\,l))}{{\mathcal{C}}^0_{\rm{SM}}(u_{\chi_u},\,l_{\chi_l})} \right] \Bigg)\,,
\end{aligned}\\
&\begin{aligned}
\Delta\rho^{d{\overline{d}}\rightarrow l {\overline{l}}}_{\rm{NLL},\Lambda } = 2 \Bigg(\rho_{{\rm{n}},\Lambda}^{d_{\chi_d} \bar{d}_{\chi_d} \rightarrow l_{\chi_l} {\overline{l}}_{\chi_l}}\mathcal{F}_D +\frac{\delta {\mathcal{C}}^0_{\rm{SM}}(d_{\chi_d},\,l_{\chi_l})}{ {\mathcal{C}}^0_{\rm{SM}}(d_{\chi_d},\,l_{\chi_l})} \sqrt{\rho_{{\rm{n}},\Lambda}^{d_{\chi_d} \bar{d}_{\chi_d} \rightarrow l_{\chi_l} {\overline{l}}_{\chi_l}}}  \,-
\\
\delta_{\chi_d,L}\delta_{\chi_l,L}\sqrt{\rho_{{\rm{n}},\Lambda}^{d_{\chi_d} \bar{d}_{\chi_d} \rightarrow l_{\chi_l} {\overline{l}}_{\chi_l}} \rho_{{\rm{c}},\Lambda}^{\phantom{A}}}\frac{g^2}{(4 \pi)^2} L_t \Re\left[ \frac{
 {\mathcal{C}}^+_{\rm{SM}}(s;\{{\rm{u}^\prime},{\rm{d}}\},\,l)) V_{{\rm{u}}^\prime{\rm{d}}}}{{\mathcal{C}}^0_{\rm{SM}}(d_{\chi_d},\,l_{\chi_l})} \right] \Bigg)\,,
\end{aligned}\\
&\begin{aligned}
\Delta\rho^{u{\overline{d}}\rightarrow \nu l^+ }_{\rm{NLL},\Lambda} = 2 &\Bigg( \rho_{{\rm{c}},\Lambda}^{\phantom{A}} \mathcal{F}_D + \frac{\delta {\mathcal{C}}^+_{\rm{SM}}(s;\{{\rm{u}},{\rm{d}}\},\,l))}{{\mathcal{C}}^+_{\rm{SM}}(s;\{{\rm{u}},{\rm{d}}\},\,l)) } \sqrt{\rho_{{\rm{c}},\Lambda}^{\phantom{A}}(s,W)} 
\\
&+ V_{{\rm{u}}{\rm{d}}}^* \frac{g^2}{(4 \pi)^2} L_u \left( \sqrt{\rho_{{\rm{c}},\Lambda}^{\phantom{A}}\rho_{{\rm{n}},\Lambda}^{u_{\chi_u} \bar{u}_{\chi_u} \rightarrow l_{\chi_l} {\overline{l}}_{\chi_l}}} \frac{{\mathcal{C}}^0_{\rm{SM}}(u_{\chi_u},\,l_{\chi_l})}{{\mathcal{C}}^+_{\rm{SM}}(s;\{{\rm{u}},{\rm{d}}\},\,l))} \right)
\\
&-  V_{{\rm{u}}{\rm{d}}}^* \frac{g^2}{(4 \pi)^2} L_t \left( \sqrt{\rho_{{\rm{c}},\Lambda}^{\phantom{A}}\rho_{{\rm{n}},\Lambda}^{d_{\chi_d} \bar{d}_{\chi_d} \rightarrow l_{\chi_l} {\overline{l}}_{\chi_l}}} \frac{{\mathcal{C}}^0_{\rm{SM}}(d_{\chi_d},\,l_{\chi_l})}{{\mathcal{C}}^+_{\rm{SM}}(s;\{{\rm{u}},{\rm{d}}\},\,l))} \right)\Bigg)\,,
\end{aligned}
\end{align*} 
where the factors $\delta {\mathcal{C}}^{0(+)}$, defined as
\begin{align*}
\delta {\mathcal{C}}^0_{\rm{SM}}&(q_{\chi_q},\,l_{\chi_l}) = \frac{\partial {\mathcal{C}}^0_{\rm{SM}}(q_{\chi_q},\,l_{\chi_l})}{\partial g^2}\delta g^2+\frac{\partial {\mathcal{C}}^0_{\rm{SM}}(q_{\chi_q},\,l_{\chi_l})}{\partial g'^2}\delta g'^2 \,,\\
&\delta {\mathcal{C}}^+_{\rm{SM}}(s;\{{\rm{u}},{\rm{d}}\},\,l)) = \frac{\partial {\mathcal{C}}^+_{\rm{SM}}(s;\{{\rm{u}},{\rm{d}}\},\,l))}{\partial g^2} \delta g^2\,,
\end{align*}
take into account the RG running of the SM couplings $g$ and $g^\prime$.

\bibliographystyle{mine}
\bibliography{bibliography}

\end{document}